\documentclass{article} 

\usepackage{setspace}

\usepackage{mathrsfs}
\usepackage{graphicx}
\usepackage{amssymb,amsmath}
\usepackage{eufrak}
\usepackage{color}
\usepackage{cite}

\usepackage{amsfonts}
\usepackage{amssymb,color}
\usepackage{amsmath,epsfig}
\usepackage{rawfonts,graphicx,amsfonts,amssymb}
\usepackage{array}
\usepackage{mathrsfs}
\usepackage{url}
\usepackage{multirow}
\usepackage{mathrsfs}
\usepackage{graphicx}
\usepackage{amssymb,amsmath}
\usepackage{eufrak}
\usepackage{color}
\usepackage{verbatim}

\usepackage{bbding,lscape}

\newcommand{\ie}{\emph{i.e.,} }
\newcommand{\eg}{\emph{e.g.,} }
\newcommand{\SINR}{\operatornamewithlimits{SINR}}
\newcommand{\Hz}{\operatornamewithlimits{Hz}}
\newcommand{\MHz}{\operatornamewithlimits{MHz}}
\newcommand{\mW}{\operatornamewithlimits{mW}}
\newcommand{\dBm}{\operatornamewithlimits{dBm}}
\newcommand{\Mbps}{\operatornamewithlimits{Mbps}}
\newcommand{\m}{\operatornamewithlimits{m}}

\newtheorem{mydef}{Definition}

\DeclareMathAlphabet{\mathpzc}{OT1}{pzc}{m}{it}

\newtheorem{theorem}{Theorem}
\newtheorem{lemma}[theorem]{Lemma}

\newtheorem{corollary}[theorem]{Corollary}

\ifodd 0
\newcommand{\rev}[1]{{\color{blue}#1}} 
\newcommand{\com}[1]{\textbf{\color{red} (COMMENT: #1) }} 

\newcommand{\response}[1]{\textbf{\color{green} (RESPONSE: #1)}}

\else
\newcommand{\rev}[1]{#1}

\newcommand{\com}[1]{}
\newcommand{\response}[1]{}
\fi

\title{Congestion Games on Weighted Directed Graphs, with Applications to Spectrum Sharing}

\author{Richard Southwell\footnote{Information Engineering Department, The Chinese University of Hong Kong,richardsouthwell254@gmail.com}, Jianwei Huang\footnote{Information Engineering Department, The Chinese University of Hong Kong,jianweihuang@gmail.com}, Biying Shou\footnote{Department of Management Sciences, City University of Hong Kong,biying.shou@cityu.edu.hk}}

\begin{document}
  \maketitle

  \begin{abstract}
With the advance of complex large-scale networks, it is becoming increasingly important to understand how selfish and spatially distributed individuals will share network resources without centralized coordinations. In this paper, we introduce the graphical congestion game with weighted edges (GCGWE) as a general theoretical model to study this problem. In GCGWE, we view the players as vertices in a weighted graph. The amount of negative impact (\eg congestion) caused by two close-by players to each other
is determined by the weight of the edge linking them. The GCGWE unifies and significantly generalizes several simpler models considered in the previous literature, and is well suited for modeling a wide range of networking scenarios. One good example is to use the GCGWE to model spectrum sharing in wireless networks, where we can properly define the edge weights and payoff functions to capture the rather complicated interference relationship between wireless nodes. By identifying which GCGWEs possess pure Nash equilibria and the very desirable finite improvement property, we gain insight into when spatially distributed wireless nodes will be able to self-organize into a mutually acceptable resource allocation. We also consider the efficiency of the pure Nash equilibria, and the computational complexity of finding them.
\end{abstract}

\section{Introduction}

Efficient resource sharing is essential in a wide range of systems in economics, engineering, {biology}, and sociology. The problem of understanding how individuals share resources in a distributed fashion is therefore scientifically important. The problem is also important in many practical situations, such as the design of communication networks, or the making {business} decisions like market entry. Interestingly, a wide range of distributed resource sharing scenarios can be modeled by congestion games. The idea is to treat individuals as players in a game, and they select which resources to use. The payoff that a player receives from using a resource is given by some decreasing function of that resource's \emph{congestion level} (\ie the total number of players using it).

Although the original congestion game (\eg \cite{rosenthal}) is \rev{rather general}, it includes no notion of space. This makes the game model unsuitable for modeling situations like spectrum allocation in wireless networks, or market entry in spatially distributed businesses. To remedy this, \cite{bilo} and \cite{ours} considered the general class of graphical congestion games. In these models, the players are represented by \emph{vertices within an undirected graph}. A player's congestion level is now the number of his neighbors who are using the same resource (\ie only linked players can congest one another). Similarly as in the original congestion game, the payoff of a player is a decreasing function of his congestion level.

There are two restrictive assumptions in the graphical congestion games models studied by \cite{bilo} and \cite{ours}. First, the congestion relationships are \emph{binary}, in the sense that  each pair of players is either linked (\ie close enough to congest each other) or not linked (\ie they never congest each other). Second, the congestion relationships are \emph{symmetric}, \ie the amount of congestion that player $n$ causes player $m$ is always equal to the amount of congestion that $m$ causes $n$.
Such assumptions can be very restrictive in many scenarios. For example, the interference caused by one wireless user to the other is a continuous function of the distance between the transmitter and receiver (rather than being binary), and the amount of negative impact that two nearby businesses in the same market have on each other may be different. This motivates us to look at a more general model, \ie  the graphical congestion game with weighted edges (GCGWE) proposed in this paper.

\begin{figure}[htb]
\centering
\includegraphics[scale=0.6]{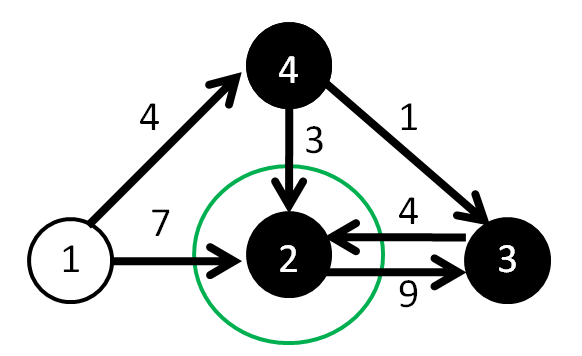}
\caption{A example of GCGWE. The players are represented by vertices in a weighted graph. The color of the vertices represent the resource that the player uses. The amount of congestion a player incurs is equal to the sum of the weights of the edges pointing to him from other \rev{players} using the same resource. For example, player $2$ incurs $3$ units of congestion from player $4$ (which is also using the \emph{black} resource) and $4$ units of congestion from player $3$, but player $2$ incurs no congestion from player $1$ (which is using the \emph{white} resource instead).}
\label{bigquasi}
\end{figure}

The GCGWE is a significant generalization of the existing graphical congestion games by considering weighted directed graphical relationship between players. In GCGWE, the amount of congestion that player $n$ causes player $m$ (when both players use the same same resource) is {represented by} a \emph{directed} edge pointing from $n$ to $m$ with a \emph{weight} $S_{n,m}$. The congestion level experienced by a player is the sum of the weights of all the edges pointing to it  from other players using the same resource. For example, in Fig. \ref{bigquasi}, player (node) $2$ uses the black resource, and experiences a congestion levee of $4+3=7$ from players 3 and 4.

The GCGWE is a powerful model, because we can select the edge weights to accurately reflect a wide range of resource allocation scenarios. For example, the players could represent business owners, the resources could represent different kinds of customers (\cite{marketshare}), and the weight $S_{n,m}$ may be a decreasing function of the distance between the business premises of players $n$ and $m$. Similarly, one can model ecosystems (\cite{eco1}) by mapping organisms to players, and different food sources which the organisms \rev{choose between} to resources.

In this paper, we will use wireless spectrum sharing to illustrate the application of GCGWEs. Wireless devices communicate through the electromagnetic spectrum. As the number of wireless devices grows rapidly, it is becoming increasingly important to understand how wireless users can share the spectrum.
Spectrum allocation has been studied extensively from a centralized point of view (\eg\cite{cent}). The complexity of the problem and the selfish nature of wireless users, however, often make it desirable to achieve efficient and fair spectrum sharing in a distributed fashion (\ie allowing users to select channels for themselves). The spectrum sharing problem has an intrinsic spatial element, because the mutual interferences generated among users on the same channel heavily depends on the locations of the corresponding transmitters and receivers. Our GCGWE model allows us to use ``weights'' to accurately reflect how much interference one user may cause to an other user. After we present the general theoretical results, we will show how spectrum sharing can be modeled accurately using GCGWEs, by choosing the correct payoff functions and edge weights.

The theoretical study of GCGWEs in this paper is two-fold. First, we want to understand \emph{how the characteristics  of player payoff functions and the underlying weighted directed graph affect the existence of a game equilibrium.} Here we will focus on examining the pure Nash equilibrium, which is a stable system state where no player can deviate from the current resource choice and improve his payoff unilaterally. Such a feature is highly desirable in many real-life resource sharing systems. Second, we shall also study \emph{the convergence of better response dynamics}, where  players asynchronously update their resource choices to improve their payoffs. This is a typical approach towards congestion game study, because asynchronous better response updates often make a good approximation of the behavior of real selfish individuals. With the answers to the above two key questions, we will understand how the qualitative features of the network topology and the payoff functions affect the ability of better response dynamics to converge to  pure Nash equilibria.


Our main results and contributions are summarized as follows:
\begin{itemize}

\item {We introduce} the GCGWE, which is a very  general model for resource allocation amongst spatially distributed individuals (Section~\ref{modelsec}).

\item {Our analytical results} reveal how the qualitative features of GCGWEs effect the players abilities to self organize into pure Nash equilibria (Section \ref{resultssec}). For some classes of GCGWE, we prove that pure Nash equilibria always exist (Theorems \ref{quasitree} and \ref{dagcomplex}). For other classes of GCGWE we prove an {even} stronger result, \ie asynchronous better response updates alway leads to a pure Nash equilibrium (Theorems \ref{twor} and \ref{sym}).

\item {We conduct extensive} simulations to verify and deepen upon our understanding about the convergence of GCGWEs to pure Nash equilibria (Section~\ref{theorysimulation}).

\item We show how the GCGWE model can be used to incorporate the realistic  physical interference model into wireless spectrum sharing. We can prescribe whether selfish user behaviors in wireless networks can lead to mutually acceptable and stable system state, based on its network topology and payoff functions (Section~\ref{wirelessmodel}).

\end{itemize}

\section{Related Work}\label{related}

The original congestion game model {was first introduced in \cite{rosenthal}}. The players select resources to use, and a player's payoff is a function of the number of players using the same resource. 
Players may use multiple resources simultaneously, and the payoffs can be \emph{resource-specific} (\ie different resources may correspond to different payoff functions).  However the payoff functions are not \emph{player-specific}, and all players are forced have the same payoff function associated with any particular resource. Every congestion game of this type has a pure Nash equilibrium. Moreover, congestion games possess the \emph{finite improvement property}, which means that better response updating will always lead to a pure Nash equilibrium in finite number of steps.

The original congestion game has been used to model a wide range of scenarios (together with other dimension of network control such as pricing, \eg \cite{Acemoglu2007,Johari2003,Ganesh2007}), and this has inspired many generalizations. \emph{Congestion games with player-specific payoff functions} were introduced in \cite{playerspec}. In these systems, different players may have different tastes for the same resource. The author restricted the discussions on \emph{singleton games}, where each player uses exactly one resource at any given time. The author showed that a singleton congestion game with player-specific payoff functions always has a pure Nash equilibrium, although there exists such games where better response dynamics can cycle and never reach a pure Nash equilibria. \emph{Weighted congestion games} (\cite{weight1,weight2}) are another generalization of the original congestion game concept. In these systems the players are associated with weights, and the congestion level experienced by a player is defined as the weighted sum of the \rev{players} using the same resources. Weighted congestion games may not possess pure a Nash equilibrium, if the players are allowed to use multiple resources simultaneously. A singleton weighted congestion game always possesses a pure Nash equilibrium, although the better response updating is not be guaranteed to converge to an equilibrium.

The graphical congestion game was originally motivated as a game of incomplete information (\cite{bilo}). More precisely, the graph represents social knowledge, so that only linked players are aware of each others resource choices. The early work on graphical congestion games focused upon a rather limited case where each player has the same linear payoff function associated with any particular resource. In \cite{ours}, we studied much more general graphical congestion game models, which allow any decreasing player-specific payoff functions. In \cite{ours} and \cite{gamenets}, we studied the singleton games on undirected graphs, and showed that that the games have the finite improvement property when there are two resources, or when the payoff functions are not resource-specific (\eg resources are homogenous). We showed that, however, there are graphical congestion games with player-specific and resource-specific payoff functions that  do not possess pure Nash equilibria.

\begin{table*}[http]
\caption{\normalsize{Properties of Several Congestion Game Models}}
\begin{center}
\scalebox{0.7}{
\renewcommand{\arraystretch}{1.0}
\begin{tabular}{ccccc}
  \hline
   & \textbf{\small{Representative}} & \textbf{\small{player- and}} & \textbf{\small{Spatial}} & \textbf{\small{Weighted}} \\
   \textbf{\small{Game type}} & \textbf{\small{literature}} & \textbf{\small{resource-}} & \textbf{\small{relationships}} & \textbf{\small{congestions}} \\
      \textbf{} &  & \textbf{\small{specific payoffs}} &  &  \\
  \hline
\small{Original congestion game} & \small{\cite{rosenthal}} &  &  &  \\
\hline
    \small{Congestion game with} & \small{\cite{playerspec}} & \small{$\checkmark$} &  &  \\
 \small{player-specific payoff} &  &  &  &  \\
 \hline
  \small{Weighted congestion game} & \small{\cite{weight2},} &  &  & \small{$\checkmark$} \\
   & \small{\cite{weight1}} &  &  &  \\
   \hline
  \small{Congestion game on} & \small{\cite{bilo}}, &  &  &   \\
  \small{unweighted undirected graph} & \small{\cite{ours}},  & \small{$\checkmark$} & \small{$\checkmark$} & \\
 & \small{\cite{gamenets}} &  &  &  \\
  \hline
  \small{\textbf{Congestion game on}} &  &  &  &  \\
  \small\textbf{{weighted directed graph}} & \textbf{\small{This paper}} & \small{$\checkmark$} & \small{$\checkmark$} & \small{$\checkmark$} \\
  \hline
\end{tabular}
}
\end{center}
\label{relatedwork}
\end{table*}

Table \ref{relatedwork} highlights the relationships between the congestion game models in the literature, and shows how the GCGWE generalizes each of these existing models. The GCGWE is most closely related to the graphical congestion game models considered in \cite{ours,gamenets}, which were also motivated by the wireless spectrum sharing applications. However, the GCGWE is much general and useful in this regard, because the freedom to choose the weighted directed edges allows us to accurately model asymmetric relationships between wireless users (see section \ref{wirelessmodel} for details). Such generalization creates many new theoretical challenges. For example, establishing the existence of pure Nash equilibria of GCGWEs with asymmetric congestion relationships requires completely new techniques. Furthermore, when proving the existence of the finite improvement property, we can no longer use the arguments from previous results that rely on the fact that congestion levels have integer values; we have to replace the arguments with new ones  that allow congestion levels to take arbitrarily small values (since edge weights and congestion levels can be arbitrarily small in GCGWEs). We will discuss the generalization in more details in Fig.~\ref{summary}, after we explained the analytical details.

\section{The Model}\label{modelsec}
\subsection{GCGWE Game Formulation}

Let us define a \textbf{graphical congestion game with weighted edges (GCGWE)} as a $5$-tuple $$( \mathcal{N},\mathcal{R},(\mathcal{R}_n)_{n \in \mathcal{N}},(f_n ^r)_{n \in \mathcal{N},r \in \mathcal{R}_n},S),$$ where
\begin{itemize}
\item $\mathcal{N} = \{ 1 , 2 ,..., N \}$ is the finite set of players. 
\item $\mathcal{R} = \{ 1 , 2 ,..., R \}$ is the finite set of resources. 
\item $\mathcal{R}_n \subseteq \mathcal{R}$ is the set of resources available to player $n \in \mathcal{N}$.
\item $f_n^r(x)$ is the payoff that player $n \in \mathcal{N}$ gets by using resource $r \in \mathcal{R}_n$, and is a strictly decreasing continuous function of the congestion level $x$ (of resource $r$ experienced by \rev{player} $n$). We will define congestion level more precisely later on.
\item $S$ is an $N \times N$ matrix of non-negative entries. Entry $S_{n,m}$ measures the amount of congestion that player $n$ causes to player $m$ when both players use the same resource. We assume\footnote{We make the assumption for notational convenience. We could relax this assumption to take into account the congestion a player causes to itself, but it seems to be easier just to modify the payoff functions to achieve this.} that $S_{n,n} = 0 $ for all $n \in \mathcal{N}$.
\end{itemize}

Matrix $S$ captures the \emph{spatial} information in the game, and can be translated into a \textbf{directed graph $D(S) = ( \mathcal{N}, \mathcal{E} )$}. The vertex set of this graph is the player set $\mathcal{N}$. An edge $(n,m)$ belongs to the directed edge set $\mathcal{E}$ if and only if the corresponding edge weight $S_{n,m} > 0$. The graph $D(S)$ describes how players can cause congestion to one another.
Sometimes we refer to a GCGWE with a spatial matrix $S$ as ``a GCGWE on graph $D(S)$''. The spatial matrix associated with the graph shown in Fig.~\ref{bigquasi} is
$$S= \left(
  \begin{array}{cccc}
    0 & 7 & 0 & 4 \\
    0 & 0 & 9 & 0 \\
    0 & 4 & 0 & 0 \\
    0 & 3 & 1 & 0 \\
  \end{array}
\right).$$

A \textbf{state} $\boldsymbol{X} = (X_1,...,X_N) \in \Pi_{n \in \mathcal{N}} \mathcal{R}_n$ represents that each \rev{player} $n\in\mathcal{N}$ picks a resource $X_{n}$ in his strategy set $\mathcal{R}_{n}$. This definition implies that we are considering singleton games in this paper, which fits into many practical applications, for example, a wireless user only has one transceiver and can only access one cellular channel.

The \textbf{congestion level} of player $n$ (when the system is in state ${ \boldsymbol X}$) is $\sum_{m \in \mathcal{N} :X_{m}=X_n}S_{m,n}$ (or simply $\sum_{m :X_{m}=X_n}S_{m,n}$). Graphically, we can think of the congestion level as the sum of the weights $S_{m,n}$ of all of the edges of $D(S)$ that are pointing to $n$, from players which are using the same resource as player $n$. In Fig.~\ref{bigquasi}, the congestion level of the player $2$ (which is using the \emph{black} resource) is $ S_{3,2} + S_{4,2} =  4 + 3 = 7$. Thus player $2$'s payoff is $f^{r} _{n} \left( \sum_{m \in \mathcal{N} :X_{m}=r}S_{m,n} \right),$ \ie  $f_2 ^{black} (7)$.


We shall begin the analysis by considering the most general form of GCGWEs, where the payoff functions are both \emph{player-specific} (\ie different players using the same resource with the same congestion level may receive different payoffs) and \emph{resource-specific} (\ie a player may receive different payoffs from different resources with the same congestion level). Later we shall will derive more properties when the payoff functions are not resource-specific.

\subsection{Better Responses, Nash Equilibria, and the Finite Improvement Property}

Consider a system in state ${ \boldsymbol X}$. Assume that a single player $n \in \mathcal{N}$ changes its resource choice to $r \in \mathcal{R}$, such that the system changes to a new state $(X_1,\cdots,X_{n-1}, r, X_{n+1}, \cdots,X_N)$.
%
%
We say that such a update is a better response update when it increases player $n$'s payoff.

\begin{mydef}
The event where a player $n \in \mathcal{N}$ changes its resource choice from $X_{n}$ to $r \in \mathcal{R}_n$ is a \textbf{better response update} if and only if
$$f_n ^r \left(\sum_{m : X_{m} = r} S_{m,n} \right) > f_n ^{X_n} \left(\sum_{m : X_{m} = X_n} S_{m,n} \right).$$
\end{mydef}

Very often we assume that  our system evolves over discrete time slots, with no more than one player updating its resource choice during any given time slot. This assumption is often used in the analysis of congestion games, and it is rather realistic when we define the time slot to be small enough such that simultaneous updating becomes unlikely.

The pure Nash equilibria are the stable resource allocations, from which no player has any incentive to deviate.
\begin{mydef}
A state ${ \boldsymbol X}$ is a \textbf{pure Nash equilibrium} if and only if no player can perform a better response update, \ie $f_n ^r \left(\sum_{m : X_{m} = r} S_{m,n} \right) \leq f_n ^{X_n} \left(\sum_{m : X_{m} = X_n} S_{m,n} \right),$
 $\forall n \in \mathcal{N}$, $\forall r \in \mathcal{R}_n$.
\end{mydef}

\begin{mydef}
A GCGWE has the \textbf{finite improvement property} if every sufficiently long sequence of better response updates leads to a pure Nash equilibrium.
\end{mydef}

\begin{figure}[t]
\centering
\includegraphics[scale=0.5]{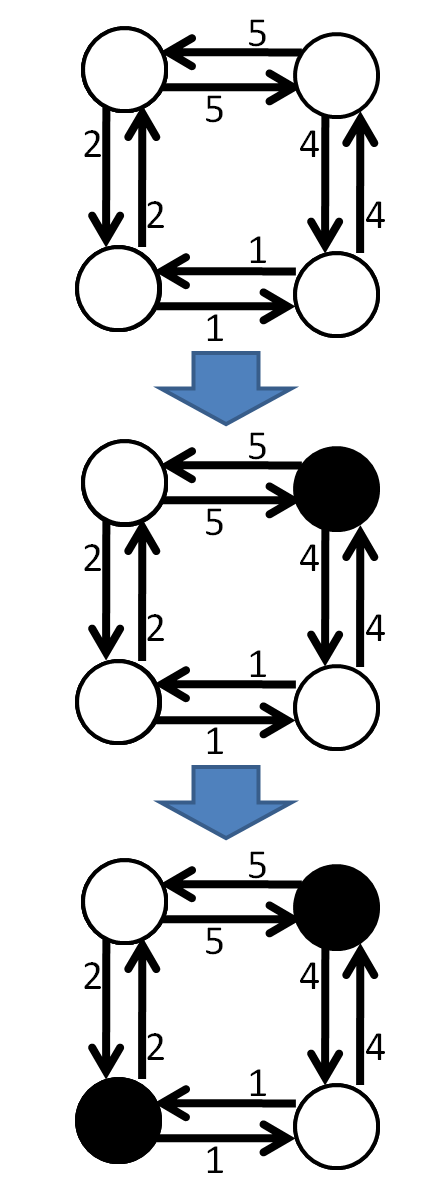}
\caption{A GCGWE with two resources (\emph{white} and \emph{black}) on an undirected graph (\ie the edges are bidirectionally symmetric), evolving under better response updates. In this case, after the player at the top right does a better response update, and then the player at the bottom left does a better response update, the system reaches a pure Nash equilibrium where where every player incurs zero congestion.}
\label{squareevolve}
\end{figure}

When a game with the finite improvement property evolves via asynchronous better response updates, it is guaranteed to reach a pure Nash equilibrium within a finite number of time slots (see Fig.~\ref{squareevolve}). Loosely speaking, this means that greedy behavior always leads to a mutually acceptable resource allocation.

%
%
%

\rev{The resources in the GCGWE represent  pure strategies that  players can choose during the game. In many games, one may further imagine  that players have the ability to use the so called \emph{mixed strategies}, where a player can use different pure strategies with different probabilities. This is beneficial mathematically, as \cite{Nash1951} shows that every game with a finite set of players and strategies (including the GCGWE) has a \emph{mixed Nash equilibrium}, where no player can increase his payoff by deviating from his current mixed strategy choice unilaterally. In our treatment of GCGWEs, we will focus on the study of pure strategy Nash equilibrium. This is because mixed strategies are often difficult to implement in practice (\cite{pricingbook}), due to reasons such as large information burden for the players (\cite{fudenberg1991,Gibbens2000}).}

\section{Results}\label{resultssec}

In this section we shall describe our main analytic results. We shall begin by discussing some examples of GCGWEs with no pure Nash equilibria. Afterwards we shall discuss various positive results about types of GCGWE that always possess pure Nash equilibria or the finite improvement property. The results in subsection~\ref{gene} are related to the most general type of GCGWEs, which have player-specific and resource-specific payoff functions. The GCGWEs we discuss in subsection \ref{resourcehomogenous} still have player-specific payoff functions, although the payoff functions are not resource-specific. We summarize our results using Table \ref{resultstab} and Fig. \ref{summary} in subsection \ref{summerhere}. Full proofs of all results can be found in the appendix.

\subsection{GCGWEs with player-specific and resource-specific payoff functions.}\label{gene}

\subsubsection{GCGWEs without pure Nash equilibria}\label{nones}

It is important to understand what kind of GCGWE do not have pure Nash equilibria, because it helps us to understand when our positive results can or cannot be generalized. Also, understanding what qualitative features induce games with no pure Nash equilibria help us to predict the situations where spatially distributed individuals cannot organize themselves into a mutually acceptable  resource allocation.


\begin{figure}[t]
\centering
\includegraphics[scale=0.45]{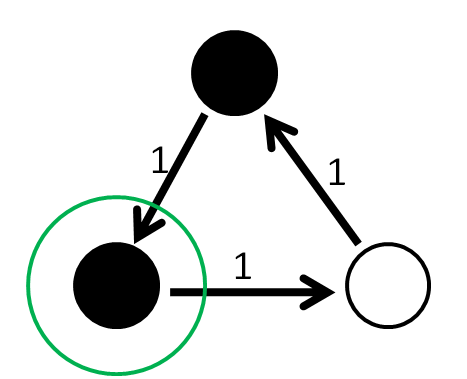}
\caption{The GCGWE that is played on a directed triangle graph (with unit weights), within which each player can choose from two resources (\emph{black} and  \emph{white}) and have homogenous payoff functions $f_n ^r (x) = -x$ (for all $n \in \{ 1 , 2 , 3 \}$ and $r \in \{ black, white \}$). This game has no pure Nash equilibrium. More specifically, no matter how the two resources are allocated, there will be some player (like the circled one above) that wishes to change their resource to avoid the congestion caused by the player linked to them.}
\label{ditri}
\end{figure}

One can construct relatively simple GCGWEs on directed graphs which have no pure Nash equilibria. An example is shown in Fig.~\ref{ditri}. If we replace the edge weights in Fig.~\ref{ditri} with any positive real numbers, the new GCGWE still has no pure Nash equilibrium. More generally, any GCGWE on a \emph{directed cycle} with an \emph{odd} number of vertices will not have a pure Nash equilibrium.

The fact that GCGWEs with no pure Nash equilibria exist is not really a new result. Our systems are generalizations of the graphical congestion games considered in \cite{ours}. In \cite{ours}, there is an example of an graphical congestion game on an undirected graph, which has no pure Nash equilibria. That game has five players, three resources, and player-specific payoff functions. We can view that example as a GCGWE with a symmetric boolean spatial matrix. What the example in Fig.~\ref{ditri} shows that, however, a GCGWE with no pure Nash equilibria can be constructed very easily if one allows asymmetric spatial matrices. However, Theorems \ref{quasitree} and \ref{dagcomplex} show that there are large classes of GCGWEs with asymmetric spatial matrices that always have pure Nash equilibria.


\subsubsection{GCGWE on a Directed Tree}\label{dirtree}

When the players are distantly scattered through space such that the graph of their congestion relationships forms a tree, a pure Nash equilibrium is guaranteed to exist.
%
\begin{mydef}
A weighted directed graph $D$ is a \textbf{directed tree} if and only if the undirected graph obtained by disregarding the directions associated with $D$'s edges\footnote{So $n$ is connected to $m$ by a single undirected edge if and only if $(n,m)$ or $(m,n)$ is an edge of $D$.} is a tree.
\end{mydef}

A directed tree is depicted in Fig.~\ref{summary}.
Every weighted tree with symmetric links is a directed tree. Also, the star and line network topologies (which are often used for the analysis of wireless networks, \eg \cite{lu2004,Sikora2006}) are examples of directed trees.

\begin{theorem}\label{quasitree}
Every GCGWE that is played upon a directed tree $D(S)$ has a pure Nash equilibrium.
\end{theorem}

\emph{Proof Sketch:} Every directed tree can be constructed by starting with a single vertex, and then adding extra vertices one at a time in such a way that every newly added vertex is linked with exactly one pre-existing vertex. When a new vertex is added to a GCGWE on a  directed tree at pure Nash equilibrium, the pre-existing vertices can adapt their strategies (taking account of the newcomers presence) to construct a new pure Nash equilibrium. We can then use induction to prove the existence of pure Nash equilibria on each  directed tree. See Appendix \ref{appendix:proofTheorem1} for a full proof. \hfill $\Box$


\subsubsection{GCGWE on a Directed Acyclic Graph}\label{isthisadaggeriseebeforeme}

\begin{mydef}
A GCGWE is \textbf{on a directed acyclic graph} when the directed graph $D(S)$ derived from the games spatial matrix $S$ contains no cycles.
\end{mydef}

\begin{theorem}\label{dagcomplex}
Every GCGWE that is played on a directed acyclic graph $D(S)$ has a pure Nash equilibrium\footnote{Note this result even holds when the payoff functions are player-specific and resource-specific.}.
\end{theorem}

\emph{Proof Sketch:} Every directed acyclic graph can be given a \emph{topological sort} \cite{topology}. A topological sort is an ordering of the vertices, such that if $i < j$ then there is no directed edge from the $j$th vertex to the $i$th vertex in the ordering. Let $k(u) \in \mathcal{N}$ denote the $u$th vertex/player which appears in some topological sort of $D(S)$. We can construct a pure Nash equilibrium of the GCGWE by having players sequentially update their strategies (in the order of $k(1),k(2),k(3), \cdots$, etc.), so that when a player updates they select the resource that maximizes their payoff against their neighbors. This leads to a pure Nash equilibrium. No player will regret their resource choice, because players updated subsequently will have no influence upon them. See Appendix \ref{appendix:proofTheroem2} for a full proof. \hfill$\Box$


We establish this result by recognizing that the resources can be allocated to individuals in an acyclic graph in such a way that the resource given to a particular individual has no effect upon the performance of an individual whose resource is allocated later. This also gives us a polynomial time algorithm for constructing a pure Nash equilibrium.

%
%

\subsubsection{GCGWE on an Undirected Graph with Two Resources}\label{undir2}


\begin{mydef}
We say a GCGWE is on \textbf{an undirected graph} when it has symmetric spatial matrix $S$, \ie  $S_{n,m} = S_{m,n},$ for all $ n,m \in \mathcal{N}$.
\end{mydef}

The following result identifies a large class of GCGWEs where better response updating is guaranteed to converge to pure Nash equilibria.

\begin{theorem}\label{twor}
Every GCGWE with $R=2$ resources on an undirected graph has the finite improvement property.
\end{theorem}

\emph{Proof Sketch:} The dynamics of GCGWEs with two resources on undirected graphs are equivalent to the dynamics of Hopfield neural networks (\cite{hop}). We exploit this fact by defining a potential function $V({ \boldsymbol X})$ on the states ${ \boldsymbol X}$ of our game, which is similar to the potential function defined for Hopfield networks. See the appendix for the definition of $V$. We show that whenever the system moves from a state ${ \boldsymbol X}$ to a state ${ \boldsymbol Y}$ because some player performs a better response update, we have $V({ \boldsymbol Y}) < V({ \boldsymbol X})$.
Since the potential function $V$ decreases with every better response update, our GCGWE cannot visit the same state more than once as it evolves via asynchronous better response updates. The number of states is finite, so the GCGWE must eventually reach a state from which no better response updates can be performed. Such a state is a pure Nash equilibrium by definition. See Appendix \ref{appendix:prooftheorem3} for the full proof. \hfill$\Box$

The assumption of having only two resources {may seem} to be restrictive at the first glance. However, Theorem~\ref{twor} has many implications, because it highlights a wide range of games (with player-specific \emph{and} resource-specific payoff functions) where players can organize themselves into pure Nash equilibria. {For example in marketing, one resource could represent entering a particular market (which would have some congestion dependent payoff), while the other resource could represent not entering the market (which would have a payoff function that is constantly zero\footnote{Technically speaking we could not include a constantly valued payoff function because our payoff functions need to be strictly decreasing, however such a function could be approximated by a function of the form $f_n^r(x) = -\epsilon.x$, where $\epsilon>0$ is sufficiently small.}). Similarly in wireless networking, one resource could represent using a particular channel, while the other represents  not using it. In scenarios where individuals decide whether or not to enter each market (channel) independently, Theorem~\ref{twor} implies the existence of pure Nash equilibria.}


\subsection{Results about GCGWEs with homogenous resources}\label{resourcehomogenous}

A GCGWE with homogenous resources is a GCGWE where all resources appear identical (from any particular player's perspective), and thus the players' payoffs are not resource-specific. There are many congestion scenarios where resources are homogenous, for example,  the assumption that resources (channels) are homogenous can be quite reasonable for some modern wireless systems (we shall discuss this more in section \ref{wirelessmodel}). Note that we still allow players' payoff functions to be player-specific in this section.

%


\begin{mydef}
A GCGWE has \textbf{homogenous resources} if and only if $f_n ^1(x)=f_n ^2(x)=...=f_n
^R(x)$, for all $ n \in \mathcal{N}$ and $ x$.
\end{mydef}

When discussing GCGWEs with homogenous resources, we will often drop the superscripts in the payoff functions (\ie writing $f_n ^r (x)$ as $f_n  (x)$).

A GCGWE becomes significantly easier to understand when resources are homogenous, because a strategy change will improve a player's payoff in this case if and only if that strategy change decreases their congestion level (see {Theorem \ref{simp}}). This fact is highly significant when the GCGWE is played on an undirected graph (see {Theorem \ref{sym}}). When a player in such a GCGWE does a better response update, they decrease their own congestion level as well as the average congestion level of their neighbors. In fact, the collective behavior of the selfish players decrease the total congestion level of the system.


\subsubsection{Better response updating is equivalent to decreasing congestion when resources are homogenous}\label{improvec}

Our next theorem states that when resource are homogenous, ``improvement'' (the increase of a player's payoff) is the same as ``decreasing congestion''.

\begin{theorem}\label{simp}
Consider a GCGWE with homogenous resources, in a state ${ \boldsymbol X}$.
The event where player $n$ changes its resource choice to $r \in \mathcal{R}_n$ is a better response update if and only if it leads to a decrease in $n$'s congestion level (\ie $ \sum_{m : X_{m} = r} S_{m,n}< \sum_{m : X_{m} = X_n} S_{m,n}$).
\end{theorem}

\emph{Proof:} The event where player $n$ changes their resource choice to $r$ is a better response update if and only if $f_n( \sum_{m : X_{m} = r} S_{m,n} ) > f_n( \sum_{m : X_{m} = X_n} S_{m,n} )$. Since $f_n$ is a strictly decreasing function, we have $f_n( \sum_{m : X_{m} = r} S_{m,n} ) > f_n( \sum_{m : X_{m} = X_n} S_{m,n} )$ if and only if $ \sum_{m : X_{m} = r} S_{m,n}  <  \sum_{m : X_{m} = X_n} S_{m,n}$. $\Box$


Theorem~\ref{simp} is fundamental and allows us to characterize more properties of GCGWEs with homogenous resources. Theorem \ref{eff} gives us an upper bound on the amount of congestion a player will incur at a pure Nash equilibrium .

\begin{theorem}\label{eff}
Suppose we have a GCGWE with homogenous resources at a pure Nash equilibrium ${ \boldsymbol X}$. Then the congestion level $\sum_{m : X_{m} = X_n} S_{m,n}$ of any player $n \in \mathcal{N}$ is no larger than ${ \left(\sum_{m=1} ^N S_{m,n}\right)}/{|\mathcal{R}_n|}$.
\end{theorem}

\emph{Proof:} We will prove the result by contradiction. Suppose to the contrary that the game is at a pure Nash equilibrium ${ \boldsymbol X}$ with $\sum_{m : X_{m} = X_n} S_{m,n} > {\left( \sum_{m=1} ^N S_{m,n}\right)}/{|\mathcal{R}_n|}$. Since $f_n$ is a strictly decreasing function, and player $n$ cannot increase their payoff by using any resource $r \in \mathcal{R}_n$, Theorem \ref{simp} implies that $\sum_{m:X_{m} = r}  S_{m,n} \geq \sum_{m : X_{m} = X_n} S_{m,n} >{\left( \sum_{m=1} ^N S_{m,n}\right)}/{|\mathcal{R}_n|}$, $\forall r \in \mathcal{R}_n$. By adding all $|R_{n}|$ inequalities together, we have $\sum_{r\in\mathcal{R}_{n}}\sum_{m:X_{m} = r}\geq |R_{n}|\sum_{m : X_{m} = X_n} S_{m,n} > \sum_{m=1} ^N S_{m,n}$. Since $\sum_{m=1} ^N S_{m,n} \geq \sum_{r\in\mathcal{R}_{n}}\sum_{m:X_{m} = r}$, we then have a contradiction of $\sum_{m=1} ^N S_{m,n} > \sum_{m=1} ^N S_{m,n}$. This proves the result. $\Box$

Theorem \ref{eff} essentially says that (at a pure Nash equilibrium) no player will have a congestion level that is above their maximum possible congestion level divided by the number of resources that are available to him. This is good news, because it means that if enough resources are available then all pure Nash equilibria will {be guaranteed to} have low levels of congestion.

\subsubsection{GCGWEs with homogenous resources on undirected graphs}\label{under}

Now we are in a position to state our central result, Theorem \ref{sym}, which identifies a large class of GCGWE's with homogenous resources which have the finite improvement property. This theorem has {very important} implications for wireless networks, as we shall discuss in Section~\ref{wirelessmodel}.


\begin{theorem}\label{sym}
Every GCGWE with homogenous resources on an undirected graph has the finite improvement property.
\end{theorem}

\emph{Proof:} Let us define the total congestion level of a state ${ \boldsymbol X}$ to be the sum of the congestion levels of all players, \ie $C( { \boldsymbol X}) = \sum_{n = 1 } ^N \sum_{m : X_n = X_{m}} S_{m,n}$.
Suppose we have a GCGWE with homogenous resources and a symmetric spatial matrix $S$, that starts in state ${ \boldsymbol X}$. Now suppose some player performs a better response update, and this converts the game state to $\boldsymbol{Y}$. This will lead to $C({ \boldsymbol Y}) < C({ \boldsymbol X})$, as shown in details in Appendix~\ref{appendix:lemma1}. To see this intuitively, note that when a player $n$ performs a better response update, it decreases their congestion level by some amount (according to Theorem \ref{simp}). Since the spatial matrix is symmetric, the \rev{sum of the congestion levels} of $n$'s neighbors decreases by the same amount, hence the total congestion level of the system decreases.

\rev{More precisely, suppose $n$'s better response update involves changing his resource from $r$ to $r'$. Now the total congestion level of the neighbors of $n$ who use $r$ will decrease, by an amount equal to the congestion level of $n$ in state ${ \boldsymbol X}$, as a result of $n$'s better response update. Also, the total  congestion levels of the neighbors of $n$ who use $r'$ will increase, by an amount equal to the congestion level of $n$ in state ${ \boldsymbol Y}$, as a result of $n$'s better response update. The congestion levels of other the players (expect for $n$ himself) will not alter as a result of $n$'s better response update. Now since $n$'s congestion level decreases as a result of the update, we have that the total congestion levels of $n$'s neighbors will decreases as a result of the update. It follows that $n$'s better response update will lead to a decrease in the total congestion level of the system.}

This implies that our GCGWE cannot visit the same state more than once when it evolves via asynchronous better response updates (because $C( { \boldsymbol X})$ decreases with every update). Since the number of states is finite, so the GCGWE must eventually reach a state from which no better response updates can be performed. Such a state is a pure Nash equilibrium by definition. \hfill $\Box$

This is an important result, because it states that when the resources are homogenous and the spatial relationships between the players are symmetric, the population will be eventually organize itself into a pure Nash equilibrium. Moreover, Theorem \ref{eff} implies that the resulting equilibria will involve relatively low levels of congestion.

\subsubsection{Computational Complexity}\label{comp}

We proved Theorem \ref{sym} by showing that the total congestion level of all the players decreases with every better response update. This fact also implies the following result about the complexity of finding efficient pure Nash equilibria.


\begin{theorem}\label{nphard}
For a GCGWE with homogenous resources on an undirected graph, it is NP hard to find the pure Nash equilibrium that maximizes the total payoff of the players among all pure Nash equilibria.
\end{theorem}

\emph{Proof Sketch:} For any undirected graph $G$, one can construct a GCGWE (with $3$ homogenous resources and $f_n^r (x) = -x$) on $G$, which has a pure Nash equilibrium under which the total payoff of the players is non-negative (\ie the highest that one can expect) if and only if $G$ can be given a proper coloring, using $3$ colors. \rev{Here a proper coloring means an assignment of one color to each vertex of the graph such that no pair of adjacent vertices share the same color.} Loosely speaking, this means finding the pure Nash equilibrium that maximizes the total payoff is at least as hard as determining whether a graph can be given a proper coloring with $3$ colors (which is an NP complete problem \cite{colorhard}). See Appendix \ref{appendix:proofTheroem7} for the full proof.\hfill $\Box$


Theorem \ref{nphard} implies that in a \emph{general} GCGWE, the problem of finding the state which maximizes the total payoff of the players will also be an NP hard problem. This is true because finding the optimal pure Nash equilibrium of a generic GCGWE is clearly at least as difficult as finding it within the special case where resources are homogenous and the graph is undirected.




 \subsection{Summary of results}\label{summerhere}

\begin{table*}[http]
\caption{\normalsize{Summary of results (Note: FIP means finite improvement property)}}
\begin{center}
\scalebox{0.7}{
\renewcommand{\arraystretch}{1.0}
    \begin{tabular}{cccccc}
    \hline
    \multicolumn{1}{c}{\textbf{\small{Network topology}}} & \multicolumn{2}{c}{\textbf{\small{Payoff Functions}}} & \multicolumn{1}{c}{\textbf{\small{Always have}}} & \multicolumn{1}{c}{\textbf{\small{Always have}}
} & \multicolumn{1}{c}{\textbf{\small{Corresponding part}}} \\
    \cline{2-3}
    \multicolumn{1}{c}{(\textbf{\small{Weighted graph type}})} & \textbf{\small{Player-}} & \textbf{\small{Resource-}} & \multicolumn{1}{c}{\textbf{\small{pure Nash}}} & \multicolumn{1}{c}{\textbf{\small{FIP?}}} & \multicolumn{1}{c}{\textbf{\small{in Section \ref{resultssec}}}} \\
        \multicolumn{1}{c}{\textbf{}} & \textbf{\small{specific?}} & \textbf{\small{specific?}} & \multicolumn{1}{c}{\textbf{\small{equilibria?}}} & \multicolumn{1}{c}{\textbf{\small{}}} & \multicolumn{1}{c}{} \\
    \hline
    \small{General directed} & \small{No}   & \small{No}    & \small{No}    & \small{No}    & \small{\ref{nones}} \\
    \small{Directed tree} & \small{Yes}  & \small{Yes}      & \small{Yes}      & \small{Unknown}      & \small{\ref{dirtree}, Theorem \ref{quasitree}} \\
    \small{Directed Acyclic} & \small{Yes}      & \small{Yes}      & \small{Yes}      &  \small{Unknown}     & \small{\ref{isthisadaggeriseebeforeme}, Theorem \ref{dagcomplex}} \\ \hline
    \small{General undirected} & \small{Yes}      & \small{Yes}      & \small{No}     & \small{No}      & \small{\ref{nones}} \\
        \small{General undirected} & \small{Yes}      & \small{No}      & \small{Yes}      & \small{Yes}      & \small{\ref{undir2}, Theorem \ref{sym}}  \\
    \small{Undirected with two resources} & \small{Yes}      & \small{Yes}      & \small{Yes}      & \small{Yes}      & \small{\ref{undir2}, Theorem \ref{twor}}  \\
    \hline
    \end{tabular}%
    }
\end{center}
\label{resultstab}
\end{table*}

\begin{figure*}[t]
\centering
\includegraphics[scale=0.30]{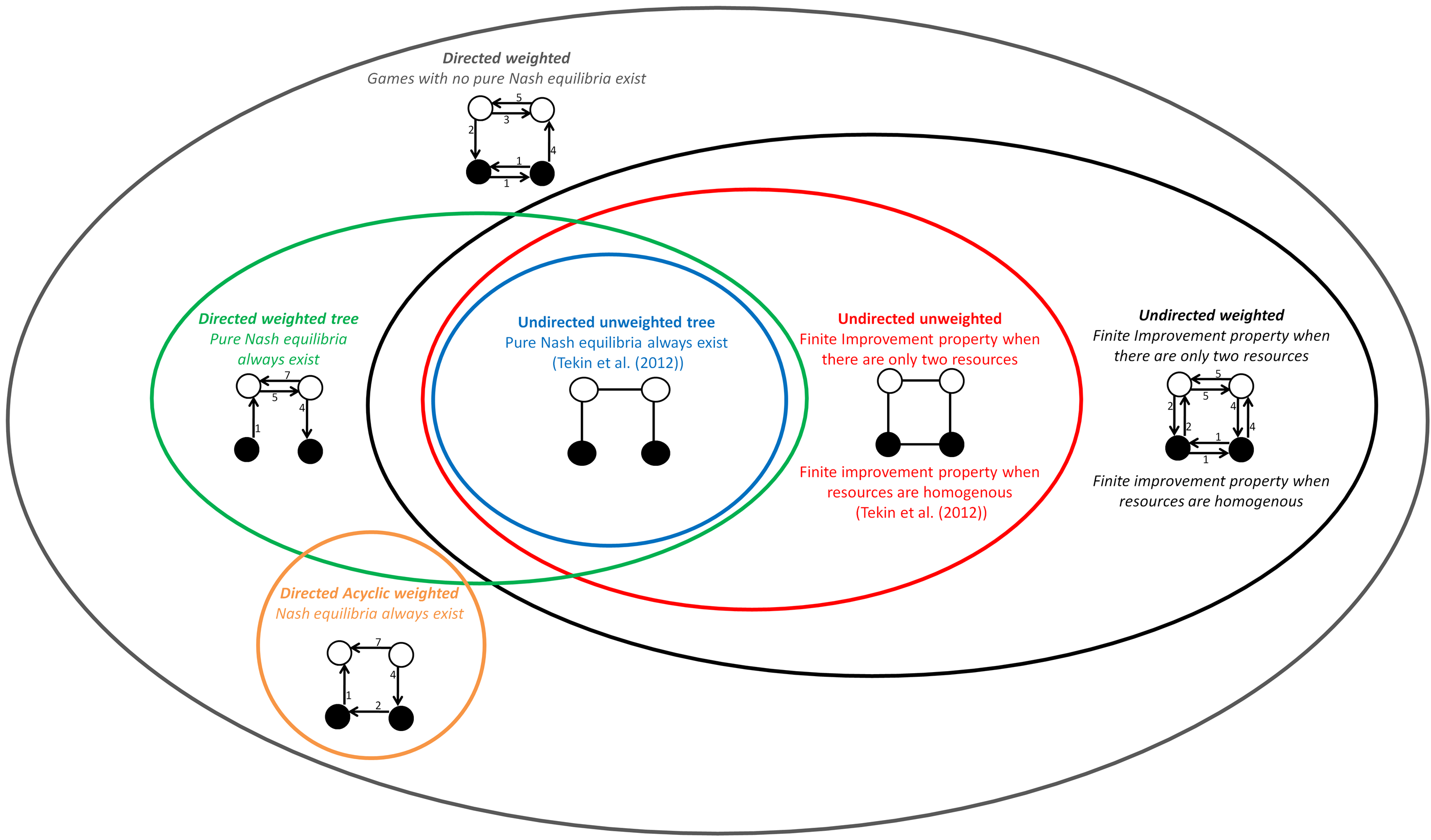}
\caption{
A Venn diagram showing the relationship between different types of GCGWEs, previously studied graphical congestion games and our main positive results. Each circle represents a class of structures, and we picture an example in each case. Our new results are \emph{italicized}. The previously known results (which are from \cite{ours}) are not italicized.}
\label{summary}
\end{figure*}

The assumptions and assertions of our results about convergence properties are summarized in Table \ref{resultstab}, which shows how different characteristics of the network topology and the payoff functions imply convergence properties. Our results show that the original congestion game of \cite{rosenthal} can be generalized a great deal, whilst preserving the finite improvement property or the existence of pure Nash equilibria. Fig.~\ref{summary} gives a visual summary of our results, and shows how they relate to previous results from \cite{ours}. We derived our positive results about the finite improvement property (\ie Theorems \ref{twor} and \ref{sym}) by substantially generalizing results in \cite{ours} from unweighted undirected graphs to weighted undirected graphs.
Furthermore, we derive completely new results on directed graph structures (Theorems \ref{quasitree} and \ref{dagcomplex}).

\section{Simulations to investigate GCGWE properties}\label{theorysimulation}

 Our analytic studies focus upon whether various kinds of GCGWEs would converge to pure Nash equilibria via better response updates. To test and expand our knowledge of this subject, we use simulations to investigate how long it will take randomly generated GCGWEs (of different kinds) to converge to pure Nash equilibria under better response updates. In particular, we investigate how the underlying graph structure, payoff functions, number of players, and number of resources, affect the convergence time of a GCGWE.

In the simulations, we consider
%
\emph{random better response update} as follows. We select a player $n$ to update, which is chosen uniformly at random, from the set of all players that \emph{can} perform a better response update. Player $n$ then choose a new resource uniformly at random from the set of resources which are better responses for him.
In terms of initialization, we will let the GCGWE start with a state where all players use the resource number one, and perform one random better response update each time slot. The simulation for one topology will stop when a pure Nash equilibria has been reached (in which case the number of time slots equals the convergence time) or until $10000$ time slots has elapsed (in which case we halt the simulation to save computer time, and because a pure Nash equilibrium may never be reached).

We perform many trials to investigate the convergence properties of our GCGWEs. Each \emph{trial} involves generating a random GCGWE by choosing its payoff functions and spatial matrix $S$, and then observing how long it will take for the system to converge to a pure Nash equilibrium under random better response updates. Suppose we have $N$ players and $R$ resources.

We consider three ways to generate random spatial matrices $S$:
\begin{itemize}
\item \emph{Random undirected graph {with uniform edge weights}}: For each $i ,j \in \{1,2,..., N \}$, if $i < j$ then we choose $S_{i,j}$ uniformly at random from $\{0,1\}$; if $i=j$ then $S_{i,j} = 0$; if $i >j$ we set $S_{i,j}$ equal to $S_{j,i}$. This method essentially generates an Erdos-Renyi random graph, where the probability that any pair of distinct vertices are linked is 1/2.
\item \emph{Random undirected weighted graph}: For each $i ,j \in \{1,2,..., N \}$, if $i < j$ then we choose $S_{i,j}$ uniformly at random from the closed unit interval $[0,1]$; if $i=j$ then $S_{i,j} = 0$; if $i >j$ we set $S_{i,j}$ equal to $S_{j,i}$. This method leads to a symmetric graph whose edges have random weights.
\item \emph{Random directed weighted graph}: For each $i ,j \in \{1,2,..., N \}$, if $i \neq j$ then we choose $S_{i,j}$ uniformly at random from the closed unit interval $[0,1]$; if $i=j$ then $S_{i,j} = 0$. This leads to a directed graph with random edge weights and may be asymmetric.
\end{itemize}

We also consider two ways to define the payoff functions:

\begin{itemize}
\item \emph{Random heterogenous payoff functions}: For each player $n$ and each resource $r$, the payoff function $f_n^r$ is a randomly selected decreasing polynomial $f_n^r (x) = -(a+bx+cx^2+dx^3)$ with coefficients $a,b,c,d$ selected uniformly at random from the open unit interval $(0,1)$.
\item \emph{Homogenous payoff functions}: For each player $n$ and each resource $r$, the payoff function $f_n^r$ has the same form $f_n^r(x) = {1}/{x}$. Under better response dynamics, any GCGWE with homogenous resources evolves in the same way as the GCGWE with homogenous payoff functions (that has the same number of resources and spatial matrix).
Recall from Theorem \ref{simp} that a better response update in a GCGWE with resource homogenous GCGWEs is equivalent of a player reduces its congestion level, and this action does not depend on the particular form of the payoff function.
\end{itemize}

\subsection{Impact of Graph Structure}

 \begin{figure}[h]
\centering
\includegraphics[scale=0.07]{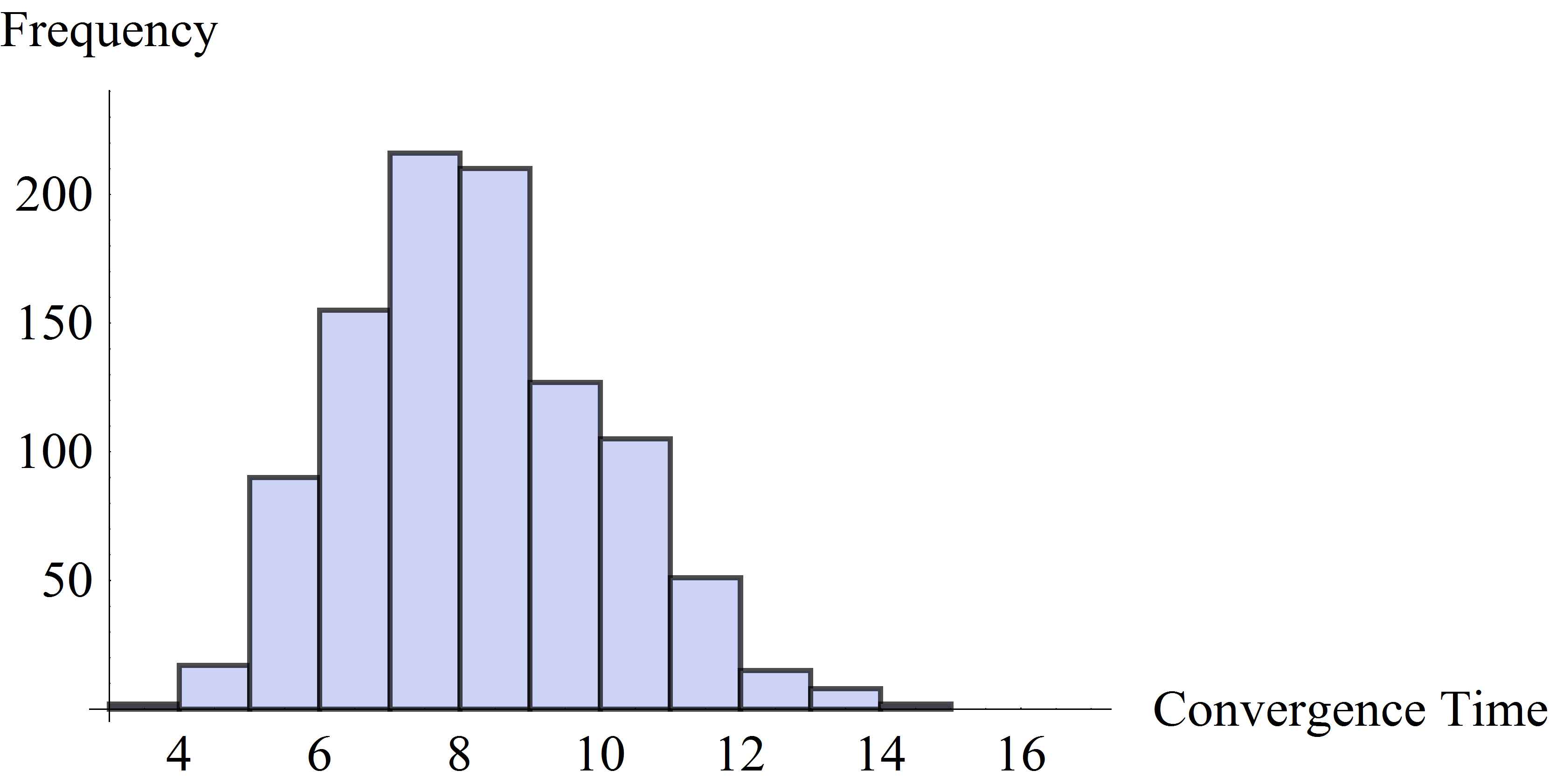}
\caption{A histogram showing the distribution of convergence times generated by $1000$ trials with $N=6$ players, $R=3$ resources, heterogenous payoff functions and random undirected graphs with uniform edge weights. Convergence occurred within every instance and the maximum convergence time is 15.}
\label{gg1}
\end{figure}

\begin{figure}[h]
\centering
\includegraphics[scale=0.07]{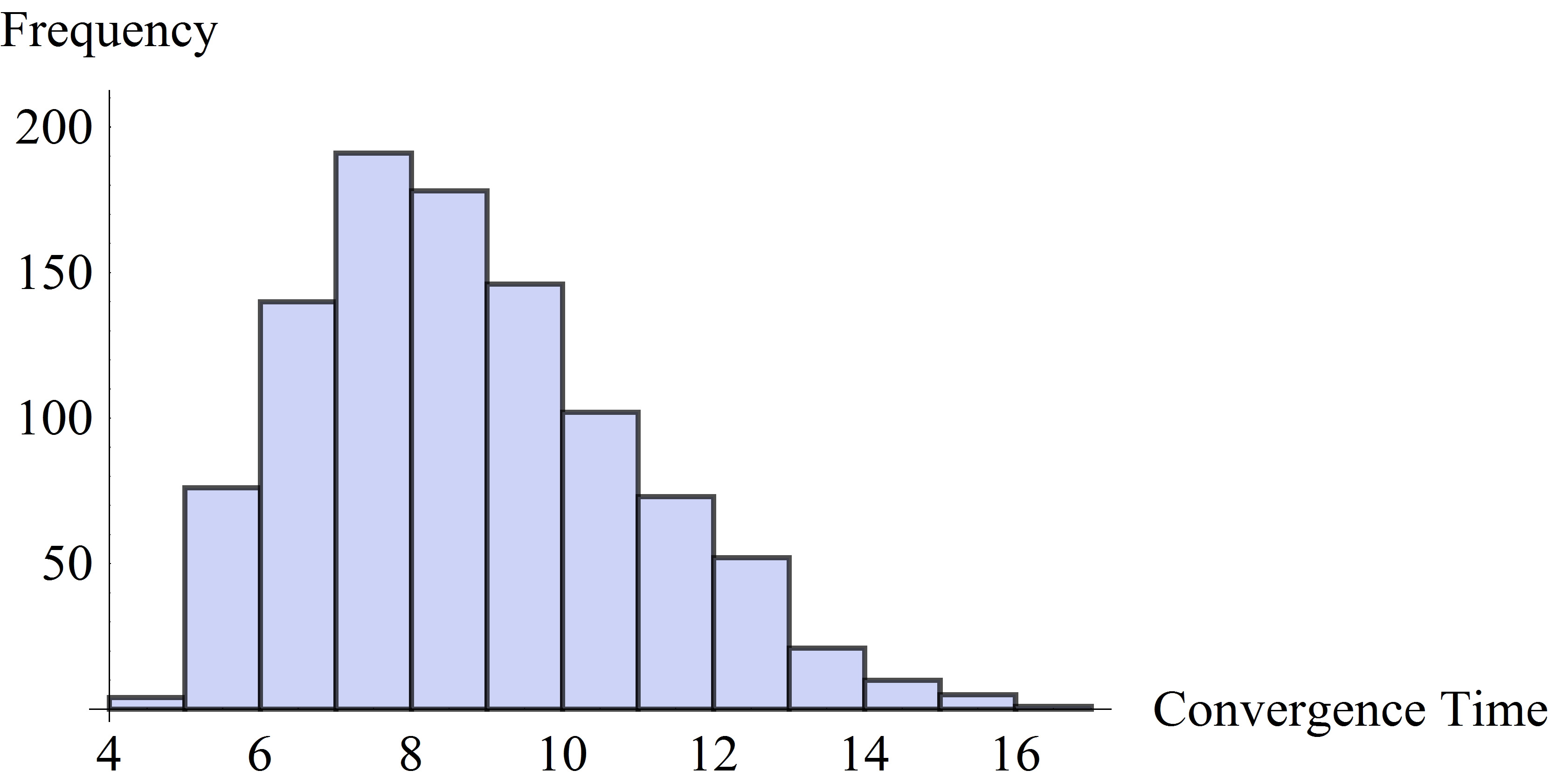}
\caption{A histogram showing the distribution of convergence times generated by $1000$ trials with $N=6$ players, $R=3$ resources, heterogenous payoff functions and random undirected weighted graphs. Convergence occurred within every instance.}
\label{gg2}
\end{figure}

\begin{figure}[h]
\centering
\includegraphics[scale=0.09]{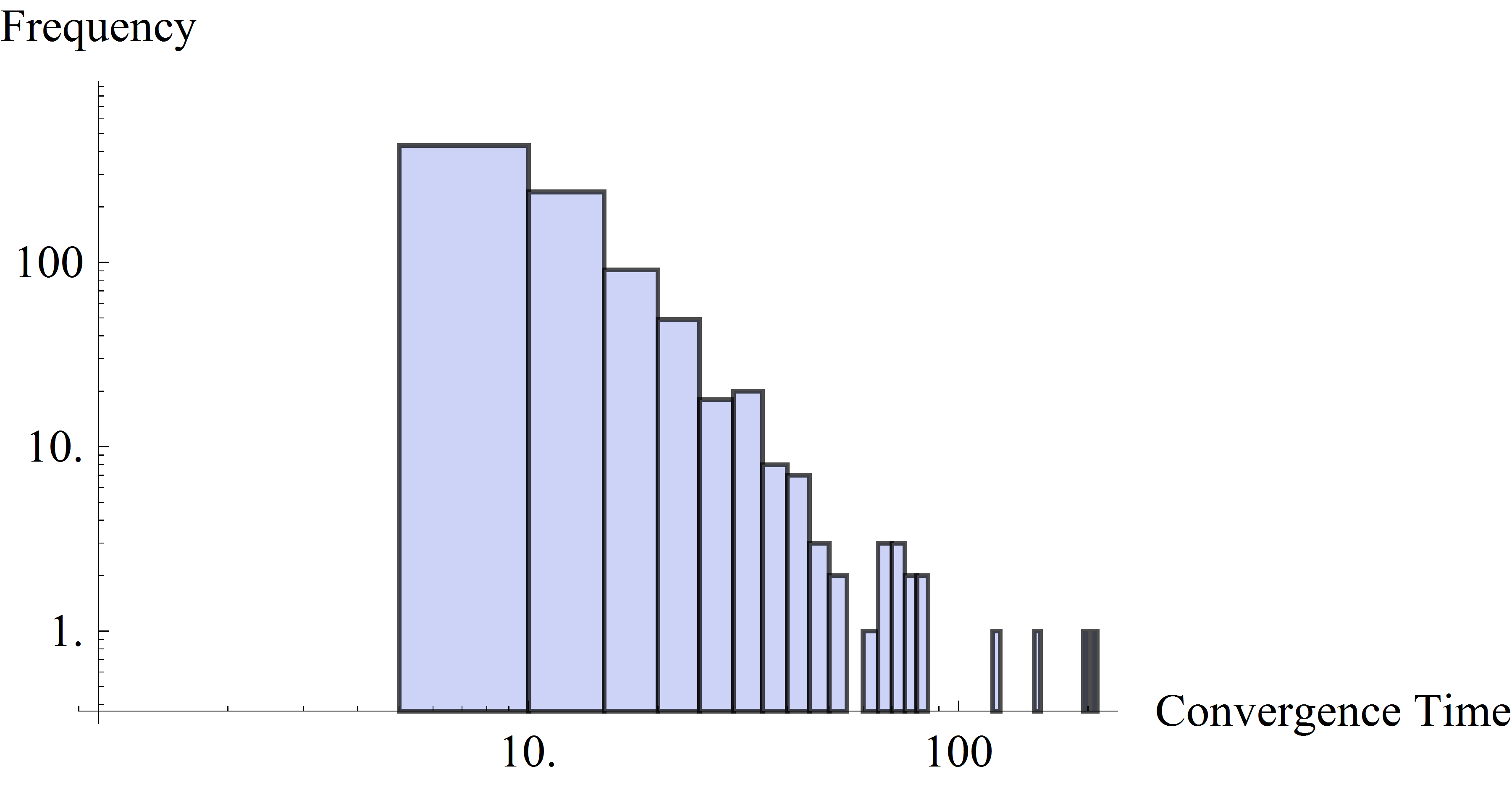}
\caption{A histogram showing the distribution of convergence times with $N=6$ players, $R=3$ resources, heterogenous payoff functions and random directed weighted graphs. To generate this distribution we performed $1000$ trials. Only $997$ of these trials converged, and we only used their convergence times to make this plot. We proved the other $3$ trials never converged. This is a long tailed distribution and so we have given the x-axis and the y-axis logarithmic scales.}
\label{gg3}
\end{figure}

We begin by investigating how the graph structure effects the convergence dynamics for random GCGWEs. We consider $N=6$ players, $R =3$ resources, and heterogenous payoff functions. Under each of the three mechanisms for randomly generating spatial matrices mentioned above, we perform $1000$ trials and observe how the convergence times of these trials are distributed.

In the first case of random undirected graphs {with uniform edge weights}, convergence occurred in each of the $1000$ trials that we conducted. Fig.~\ref{gg1} shows how the systems converge quickly (within $15$ time slots) in each case.
When we consider the second case of random undirected weighted graphs, Fig.~\ref{gg2} illustrates that the convergence times are very similar as that in Fig.~\ref{gg1}.
Again, convergence occurred within each of our $1000$ trials. However, as we note in subsubsection \ref{nones}, it is possible to construct graphs in this category such that there is no Nash equilibria. The simulation imply that such cases are rare in practice.


\begin{figure}[h]
\centering
\includegraphics[scale=0.07]{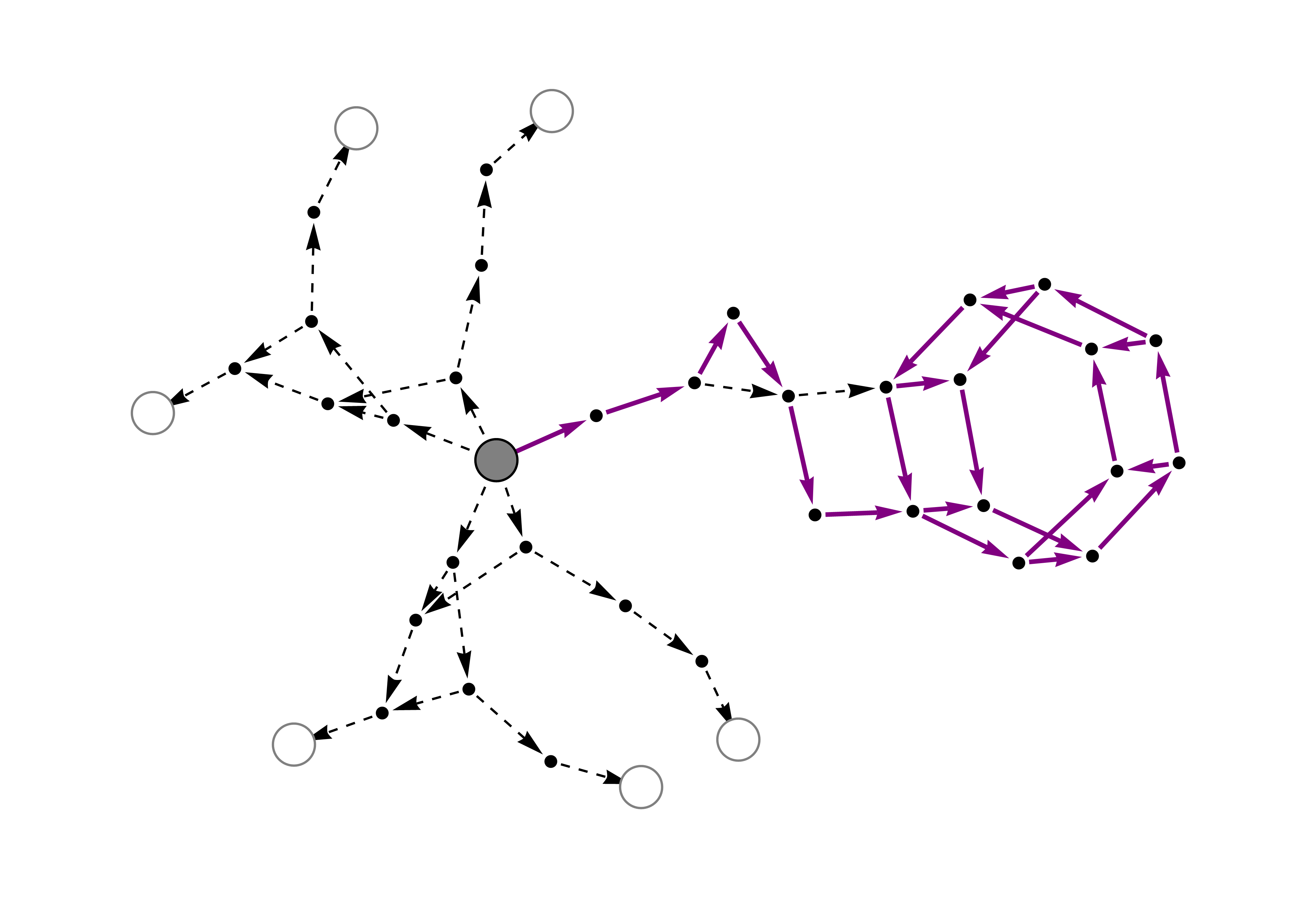}
\caption{A fragment of the state space associated with one of the none convergent simulations encountered in the generation of Fig.~\ref{gg3}. Each point represents a state (a way to associate players with resources), each arrow represents a state transition that can occur because a player does a better response update. The large gray node represents the initial condition (where all players are using resource $1$). The white nodes are the pure Nash equilibria of the system. The path through the state space the system actually took during the simulation, is shown by the purple arrows. Note that the state gets trapped in a piece of the state space which is not a pure Nash equilibrium, and from which it cannot escape.}
\label{hh5}
\end{figure}

When we consider the third case of random directed weighted graphs (\ie asymmetric relationship among players), non-convergent systems become more common. Even for those that converge, they may take much longer times. Fig.~\ref{gg3} illustrates this phenomenon, where $997$ out of $1000$ trials converged within $10000$ time slots. We studied the remaining three non-convergent cases in more details. In two of these three cases, convergence did not occur because pure Nash equilibria do not exist (meaning that convergence is completely impossible, from any initial condition). The last case is more interesting, and the system state transition is shown in Fig.~\ref{hh5}. In this case, it is possible for the system to converge to a pure Nash equilibrium because it does exist, and we can find such paths that lead to a pure Nash equilibrium. However, our random better response updates were stuck in a recurrent part of the state space (the purple part) and thus did not converge. To put it more whimsically, the system had a chance to get to a pure Nash equilibrium, but it was unlucky, and ended up falling into an inescapable hole.

In addition to the fact that convergence is not guaranteed when the spatial matrix is asymmetric, Fig.~\ref{gg3} also reveals that the distribution of convergence times looks different to the previous cases in Figs.~\ref{gg1} and \ref{gg2}. The distribution in Fig.~\ref{gg3} is \emph{long tailed}, meaning that very long convergence times can occur with non-negligible probabilities. This reveals an intrinsic difficulty with studying convergence times through simulations. It is possible for a system to take a very long time to converge, and it is also possible that the system will never converge. In the systems we study there are only $729$ states, and so it is possible to make a complete picture of the state space and rigorously verify whether pure Nash equilibria exist and whether they can be reached from the initial condition (as in Fig.~\ref{hh5}). However, in non-convergent simulations involving more resources and players, it can be very difficult to know whether one has run the system for long enough, or whether convergence is will never occur.

\subsection{Impact of Resources Homogeneity}

\begin{figure}[h]
\centering
\includegraphics[scale=0.07]{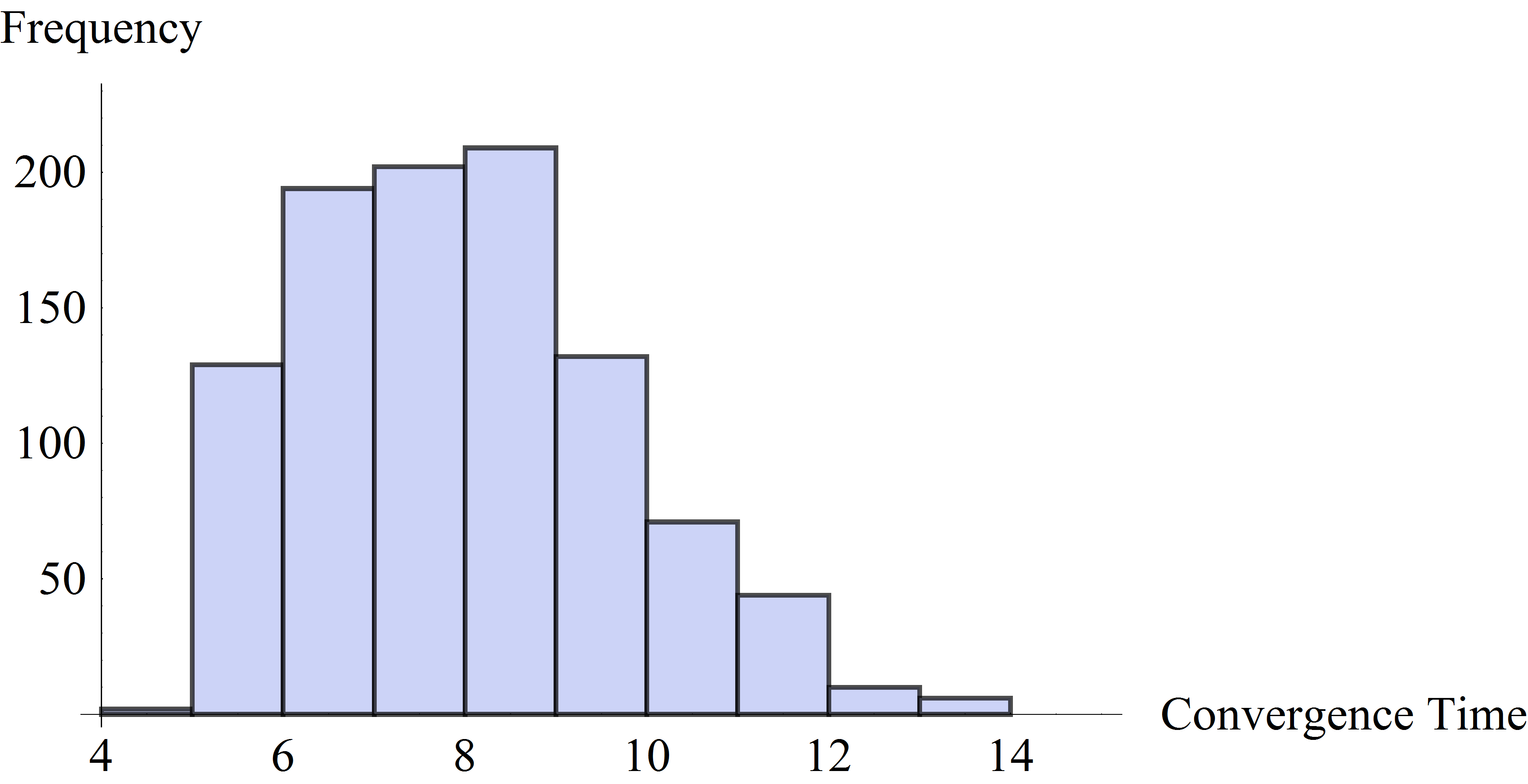}
\caption{A histogram showing the distribution of convergence times generated by $1000$ trials with $N=6$ players, $R=3$ resources, homogenous payoff functions and random undirected weighted graphs. Convergence occurred within every instance.}
\label{hh2}
\end{figure}

When the resources are homogenous, convergence is guaranteed in special cases. In Fig.~\ref{hh2} we show how the convergence times are distributed for random GCGWEs with $N=6$ players, $R=3$ resources, random undirected unweighted graphs and homogenous payoff functions. Each of the $1000$ trials converged. This agrees with Theorem \ref{sym}, which implies that these kind of systems will always converge to a pure Nash equilibrium eventually. Also, Fig.~\ref{hh2} reveals the pleasing fact that this convergence seems to occur very quickly in practice. The mean convergence time of these trials was $7.531$, which is less than the mean convergence time of the case where the payoff functions were heterogenous (\ie Fig.~\ref{gg2}), which was $8.091$.

\subsection{Impact of Numbers of Players and Resources}

\begin{figure}[h]
\centering
\includegraphics[scale=0.6]{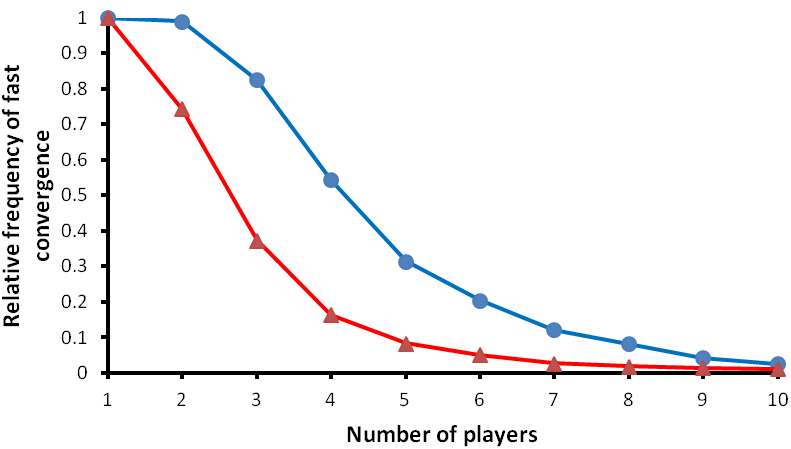}
\caption{The relative frequency of fast convergence, for GCGWEs with $R=3$ resources, and heterogenous payoff functions. The x-axis shows the number of players. The y-axis shows the fraction of $1000$ trials which converge to pure Nash equilibria within ten time slots. The blue circular points were generated using GCGWEs on random undirected weighted graphs, the red triangular points were generated using GCGWEs on random directed weighted graphs.}
\label{playersvsfreq}
\end{figure}

\begin{figure}[h]
\centering
\includegraphics[scale=0.6]{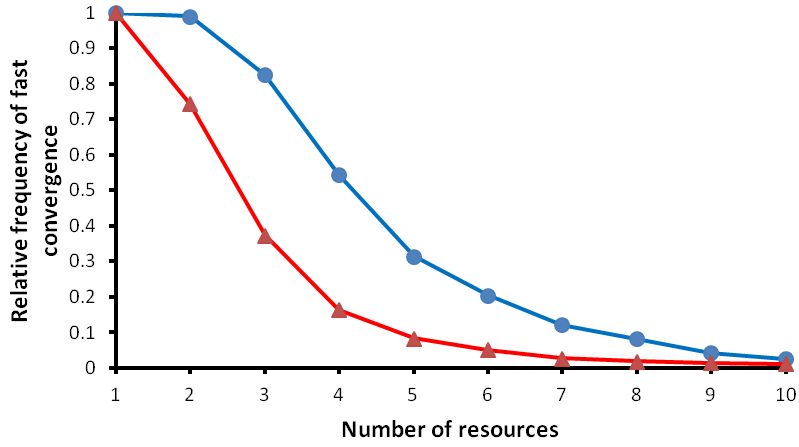}
\caption{The relative frequency of fast convergence, for GCGWEs with $N=6$ players, and heterogenous payoff functions. The x-axis shows the number of resources. The y-axis shows the fraction of $1000$ trials which converge to pure Nash equilibria within ten time slots. The blue circular points were generated using GCGWEs on random undirected weighted graphs, the red triangular points were generated using GCGWEs on random directed weighted graphs.}
\label{resourcesvsfreq}
\end{figure}

It is difficult to compare the convergence times of different kinds of GCGWEs directly. This is because the notion of \emph{expected convergence time} is undefined when the system may not converge, and the variance of times taken in convergent systems may be very large (as illustrated in Fig.~\ref{gg3}). For these reasons, we compare different kinds of GCGWEs using the \emph{relative frequency of fast convergence}. We define this to be the fraction of our $1000$ trials which converge to a pure Nash equilibrium within ten time slots. Fig.~\ref{playersvsfreq} we show how the relative frequency of fast convergence depends upon the number of players. This figure shows how the relative frequency of fast convergence decreases gradually with the number of players, and GCGWEs with random undirected weighted graphs are more likely to converge within ten time slots than GCGWEs with random directed weighted graphs. In a similar way, Fig.~\ref{resourcesvsfreq} shows how the relative frequency of fast convergence depends upon the number of resources (when the number of players is six).

\section{Modeling Wireless Networks}\label{wirelessmodel}

The spectrum sharing problem has an intrinsic spatial element, because the mutual interferences generated among users on the same channel heavily depends on the locations of the corresponding transmitters and receivers. The GCGWE allows us to use ``weights'' to reflect how much interference one user may cause to an other user. By selecting edge weights appropriately, we can account for the spatial aspects of spectrum sharing in an accurate way. This is {very useful because} the spatial aspect of spectrum sharing is less understood than many other aspects (\cite{geo}). In this section we shall describe how the GCGWE can be used to model spectrum sharing, in accordance with the physical interference model (\eg \cite{physicalinterference}). And then we shall use the GCGWE to simulate spectrum sharing.

Many game theoretic models have been used to study spectrum sharing (\eg \cite{requiresinfo,wu2009repeated,marketp,Huang:2006p1625}), but these generally assume that the users have complete knowledge about the network parameters and each other's information. Congestion game based models have the advantage of modeling network scenarios where the users have limited information, which is the case in many networks (cognitive radio networks in particular). In \cite{ours} and \cite{gamenets}, we considered how congestion games on undirected, unweighted graphs could be used to model spectrum sharing (with players representing wireless users and resources representing channels). Using an undirected unweighted graph to represent interference relationships in this way corresponds to the protocol interference model in \cite{gupta2000capacity}, within which a pair of users are either considered \emph{linked} (in which case they can cause one another some fixed amount of interference) or \emph{not linked} (in which case they are considered to be too distant to interfere with one another).

Graphical congestion games serves as more realistic models for spectrum sharing than classical congestion games, because they have player-specific payoff functions and allow for \emph{spectrum reuse} (where distantly spaced individuals can use the same channel without mutual interference). However, the protocol inference model does not fully
capture the real interference relationships between the users. In reality, inference level depends on transmission power levels  and decreases continuously with separation distance. Interference effects can be captured much more accurately using the signal-to-noise ratio (SINR) model (\eg \cite{physicalinterference}). Our models in this paper are powerful enough to incorporate the SINR-based physical interference model, and thus are generalizations of the models introduced in \cite{ours}.

\subsection{Modeling Physical Interference Model with Fixed Transmission Power}\label{SINRcor}
Consider a wireless network where each user is a fixed transmitter-receiver pair. We model the interference by the physical interference model, where  the interference received by a user is the summation of the power received from all other users in the network.
%
 The maximum achievable transmission rate (according to the Shannon capacity) that a user $n$ gets by using channel $r$ is $B_r \log _2 \left( 1 + \SINR \right)$, where $B_{r}$ is the bandwidth of channel $r$ and $\SINR$ is the signal-to-interference plus noise ratio,
$$\SINR = \frac{h_{n,n} P_n}{\tau_0 B_i + \sum_{m: m \neq n, X_{m} = r} h_{m,n} P_{m}}.$$
%
Here $\tau_0$ is the thermal noise density, $P_{m}$ is the transmission power of $m$'s transmitter, and $h_{m,n}$ is the channel gain from $m$'s transmitter to $n$'s receiver.

If each user has a fixed transmission power (which is the default operation mode in today's Wi-Fi networks), then the spectrum sharing scenario can be modeled as a GCGWE $( \mathcal{N},\mathcal{R},(\mathcal{R}_n)_{n \in \mathcal{N}},(f_n ^r)_{n \in \mathcal{N},r \in \mathcal{R}_n},S)$, where each player $n \in \mathcal{N}$ corresponds to a fixed transmitter-receiver pair. Each resource $r \in \mathcal{R}$ corresponds to an orthogonal channel. When channels have an equal bandwidth (which is true in Wi-Fi , WiMax,  and LTE networks) and the channels are interleaved (and thus have the same channel conditions for the same user), the system corresponds to a GCGWE with homogenous resources.
Each user $n$ has a \emph{user-dependent} available channel set $\mathcal{R}_n \subseteq \mathcal{R}$. This flexibility is especially useful for modeling cognitive radio networks, where the channels available to a secondary user depend on the activities of the licensed users within its vicinity. Each player uses exactly one resource/channel at a given time, due to limitation of the hardware. $S_{m,n}$ measures the amount of interference that $m$ causes $n$ when both users are on the same channel. More precisely, $S_{n,n} = 0$ and $S_{m,n} = h_{m,n}P_{m}$ for $n \neq m$. Player $n$'s payoff of using resource $r$ depends on the interference level $x = \sum_{m : X_{m} = X_n} S_{m,n}$, and is equal to
\begin{equation}%
f_{n} ^r (x) = B_r \log _2 \left( 1 + \frac{h_{n,n}P_{n}}{ \tau_0 B_r + x } \right). %
\label{sinreqn}
\end{equation}

When the power levels of the users are equal, and distinct transmitter-receiver pairs are distantly spaced (relative to the distance between individual transmitters and their receivers), the assumption that the interference relationship between users is symmetric (\ie $h_{n,m} = h_{m,n}$) is at least approximately valid.
Such cases correspond to GCGWEs with symmetric spatial matrices, or ``on undirected graphs''.


\subsection{More General Modeling of Wireless Networks}

The GCGWE can model much more general wireless communication problems than the one described above. In particular, the models can include:
\begin{itemize}
\item \emph{User-specific transmission technologies:} Users may have different payoffs and different channel preferences because of different transmission technologies. We can choose player-specific payoff functions (instead of the same Shannon capacity)  to model this.
\item \emph{User-priorities:} We can also use edge weights to reflect  different user priorities  in cognitive radio networks. For example, if $n$ is a primary license holder and $m$ is an secondary unlicensed user, then we could set $S_{n,m}$ to be very large to reflect the price that $m$ must pay (or the punishment which $m$ may receive) for causing interference to the license holder $n$.
\end{itemize}

{Based on the analytic results shown in Section \ref{resultssec}, we can prescribe whether selfish user behaviors in wireless networks can lead to mutually acceptable and stable system state, based on its network topology and payoff functions.}

\subsection{Simulating wireless networks}\label{simulations}
 The fact that GCGWEs are general enough to emulate the SNIR model gives us the valuable opportunity to study a realistic model of spectrum sharing. In this section, we will simulate spectrum sharing in wireless networks using GCGWEs. In particular, we will investigate how $N=20$ selfish radio users (scattered across a square region of length $L$) will share $R=5$ homogenous channels. We study how the users' ability to share the spectrum is influenced by $L$. We suppose that each player corresponds to a fixed transmitter-receiver pair that wishes to maximize its transmission rate by selecting the best channel, in the same way as we described in Section \ref{SINRcor}. We shall also make the following assumptions:
\begin{itemize}
\item Each of user $n$ transmits at a fixed power level of $P_n = 100 \mW$.
\item Each channel $r$ has a bandwidth of $B_r = 20 \MHz$, and is available to every user.
\item The payoff that a user $n$ gets for using a channel $X_n = r$ is equal to its transmission rate, as given by Equation (\ref{sinreqn}).
\item We shall use the distance-based physical interference model (\cite{physicalinterference}), by writing the channel gain $h_{m,n}$, from user $m$'s transmitter to $n$'s receiver (see Equation (\ref{sinreqn})) as $h_{m,n} = 1/d_{m,n}^{\alpha}$, where $\alpha$ is the attenuation factor and $d_{m,n}$ is the distance from $m$'s transmitter to $n$'s receiver.
\item We will suppose that the attenuation factor $\alpha = 4$ and the spectral noise density $\tau_0 = -174 \dBm/\Hz = 10^{-17.4} \mW/\Hz$.
\item We place each transmitter at a point (chosen uniformly at random) from our $L \times L$ square.
    Each receiver is uniformly randomly located within $100\m$ of its transmitter. We insure that no receiver is with $1\m$ of a transmitter (since our distance based SINR model breaks down at such close ranges).
\end{itemize}

\begin{figure}[h]
\centering
\includegraphics[scale=0.35]{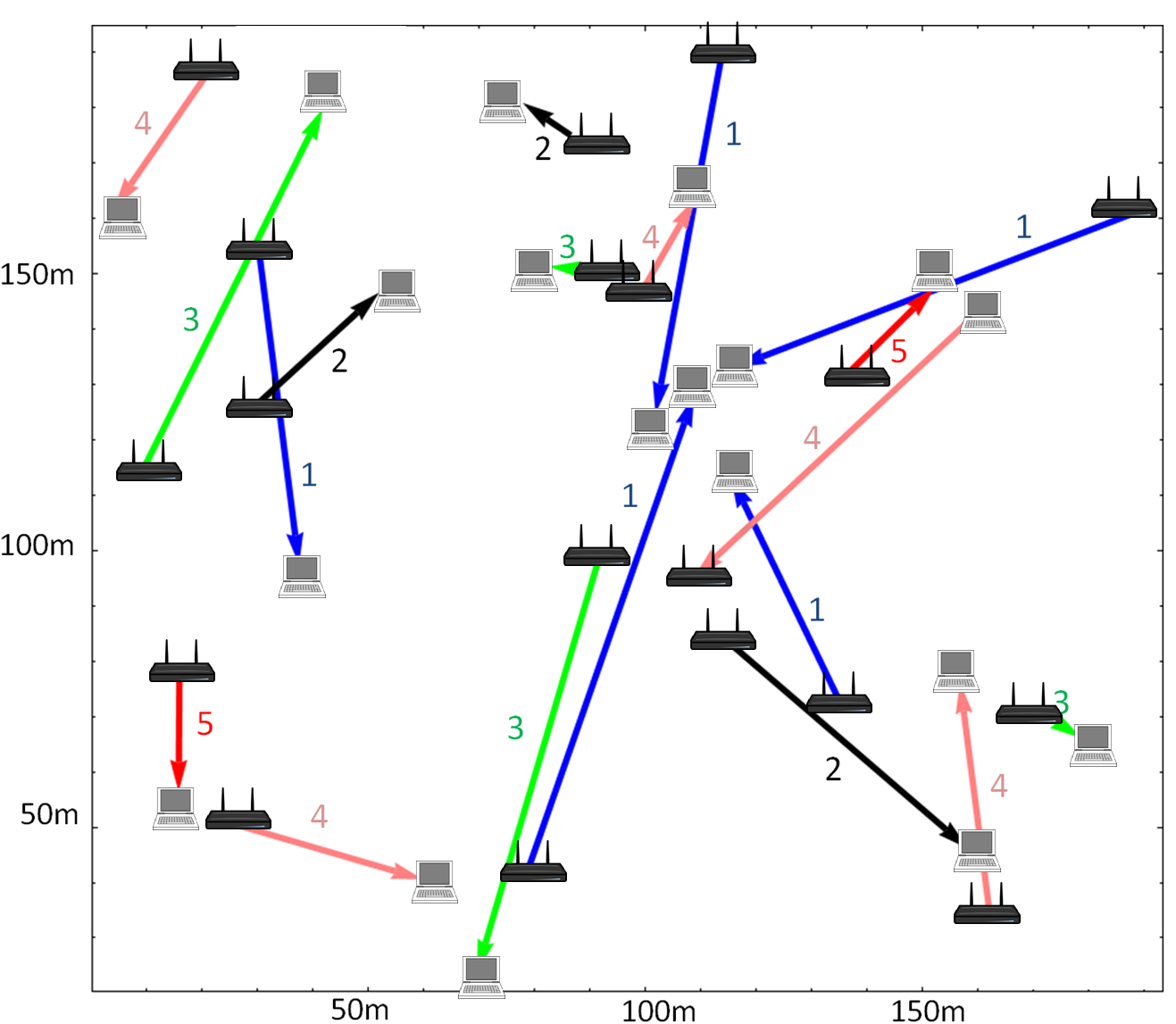}
\caption{A pure Nash equilibrium in a network where the $N=20$ users are scattered across a square of length $L = 200 \m$. Each arrow points from a transmitter (represented by a black router) to its receiver (represented by a gray computer). The link colors represent the channels that users choose at this pure Nash equilibrium. The average transmission rate of users at this pure Nash equilibrium is $101 \Mbps$. Notice how links (users) of the same color naturally spread out to avoid strong mutual interferences.}
\label{scatter}
\end{figure}

For each simulation run, we randomly generate a network and randomly allocate one of the five channels to each user. Then the network evolves under random better response updates ({as described} in Section~\ref{theorysimulation}) until a pure Nash equilibrium has been reached (or some pre-specified large number of time slots have elapsed). We show the choices of channels in a pure Nash equilibrium of a network under our parameters in Fig.~\ref{scatter}.
\begin{figure}[h]
\centering
\includegraphics[scale=0.45]{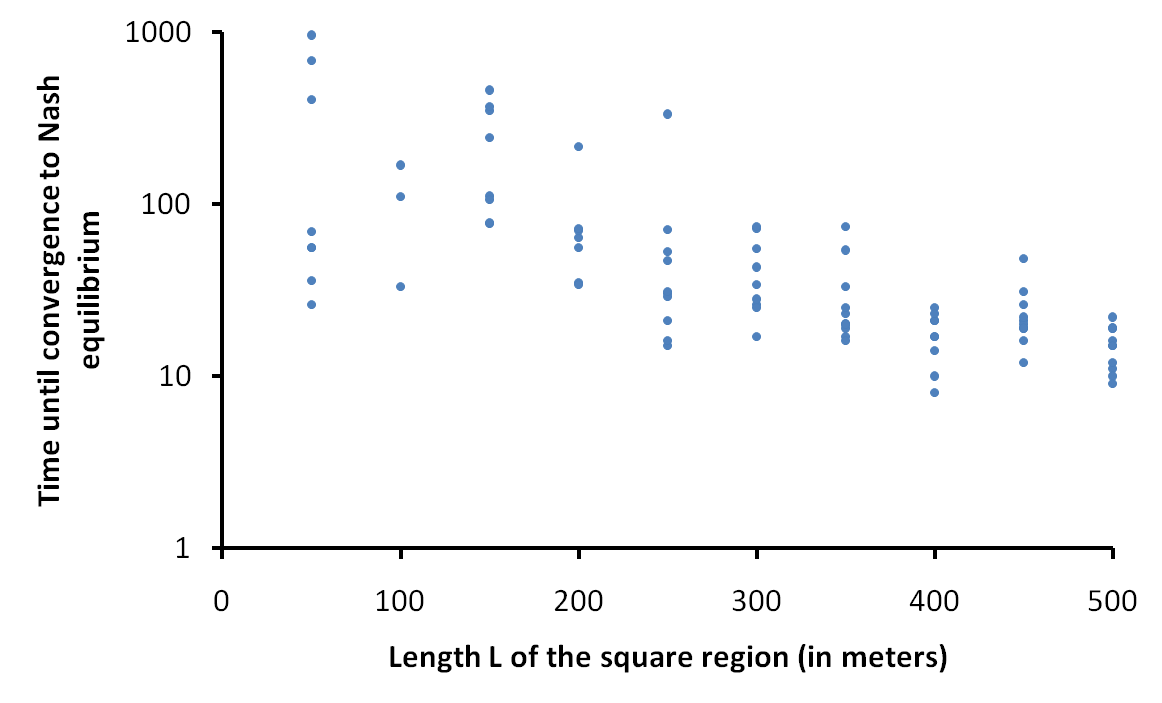}
\caption{The $x$-axis gives the length $L$ (in meters) of the square region which the users are scattered over. We performed $10$ simulations for each $L \in \{ 50\m , 100\m,..., 500\m \}$. We ran each simulation for up to $2000$ time slots. The $y$-axis shows how many time slots it takes a given simulation to converge. We do not show the points corresponding to simulations which did not converge. Note that the scale of the $y$ axis is logarithmic.}
\label{timescatter}
\end{figure}
\begin{figure}[h]
\centering
\includegraphics[scale=0.80]{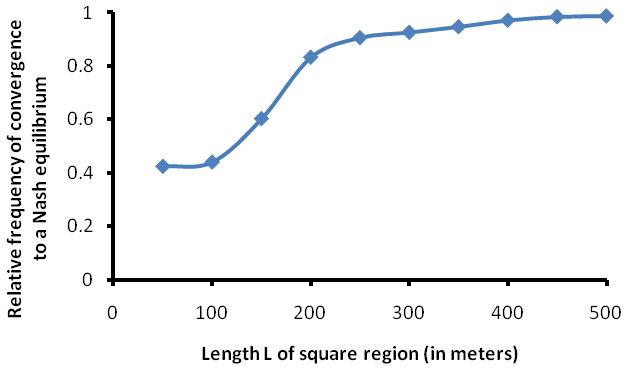}
\caption{The $x$-axis gives the length $L$, in meters, of the square region which the users are scattered across. We performed $1000$ simulations for each value of $L$. The $y$-axis gives the fraction of these simulations which converged to a pure Nash equilibrium within $500$ time slots.}
\label{lengthconvergenceprob}
\end{figure}


The majority of our simulation runs converged to pure Nash equilibria. As Fig.~\ref{timescatter} shows, however, convergence time is highly variable in simulation runs.  Figure \ref{timescatter} shows that $20$ users normally reach a pure Nash equilibrium within $200$ time slots. In rare cases, however, it can take over $1000$ time slots for a simulation to converge. In this regard, we re-encounter the problem we saw in Section~\ref{theorysimulation}: sometimes a system will never converge, and sometimes it will take a very long time to converge. Also, unlike in Section~\ref{theorysimulation}, each of these systems has $5^{20}$ states. This make it impractical to rigorously verify whether a given system has the finite improvement property or pure Nash equilibria.

For this reason, we studied how the relative frequency that a given system will converge within $500$ time slots, depends upon the geometry of the network. Figure \ref{lengthconvergenceprob} shows that it is easier for the users to organize themselves into a pure Nash equilibrium when the area they are spread across is larger. This effect agrees with our findings from section \ref{resourcehomogenous}. When the users are scattered across a large area, the distance between distinct transmitter-receiver pairs will often be much larger than the distance between a particular transmitter and its receiver. This will cause the interference relationship between the users to be approximately symmetric (in that the distance from $n$'s transmitter to $m$'s receiver will be approximately equal to the distance from $m$'s transmitter to $n$'s receiver). This means the system will (approximately) correspond to a GCGWE with homogenous resources and a symmetric spatial matrix. In this case, Theorem \ref{sym} states that the system will have the finite improvement property, and therefor will eventually converge to a pure Nash equilibrium. 



\section{Conclusion}\label{discussion}



In this paper, we introduce the graphical congestion game with
weighted edges (GCGWE), which is a general model for studying how selfish and
spatially distributed individuals will share resources. Although we use spectrum sharing to illustrate the effectiveness of GCGWE, such a model can really be used in many different network scenarios. By identifying which GCGWEs
possess pure Nash equilibria and the finite improvement property,
we gain insight into when spatially distributed wireless nodes
will be able to self organize into a mutually acceptable resource
allocation. We also consider the efficiency of the pure Nash
equilibria, and the computational complexity of finding them.

Our results and simulations suggest that the topology of the network is more important than the forms of the payoff functions, with respect to the ability of the players to self organize to pure Nash equilibria. {One of our key results}, Theorem \ref{sym}, states that any GCGWE with symmetric congestion relationships and homogenous resources has the finite improvement property. In addition to our observations about the positive effects of symmetric congestion relationships, we have also observed the negative effects of asymmetric congestion relationships. In particular, we found that GCGWEs on directed graphs with no pure Nash equilibria can easily be constructed, and simulations reveal that some GCGWEs on random directed graphs never (or took a very long time to) converge to a pure Nash equilibrium under better response updates.

The fact that convergence properties are so sensitive to the qualitative features of the network the game is played on, validates our approach, because it implies that an accurate model of the spatial relationships between players (such as the GCGWE, as opposed to the standard ``congestion game on an unweighted undirected graph'') is necessary in order to predict the behaviour of spatially distributed, resource sharing individuals. In the future, we shall consider GCGWEs where players can use multiple resources simultaneously. We shall also attempt a finer characterization of which GCGWEs do not have pure Nash equilibria.

\section{Appendix}

In this section we give full proofs to some of the results that we only mentioned ``proof sketches'' of in the main paper.

\subsection{Proof of Theorem \ref{quasitree}}\label{appendix:proofTheorem1}

The result clearly holds for any 1 player GCGWE. We will prove by induction.

Suppose every game on a quasi-tree with $N$ vertices has a Nash equilibrium. Now suppose $\psi$ is a GCGWE on a quasi-tree $T$ with $N+1$ vertices. Now $T$ can be constructed by taking a quasi-tree $T'$, on $N$ vertices, and then adding a vertex $N+1$ together with a (one or two-way) link to some vertex $n$ of $T'$. Let us consider the game, restricted to the $N$ vertices of $T'$. Now, since this is a GCGWE on an $N$ vertex quasi-tree, it has a Nash equilibrium ${ \boldsymbol X}$. Now let us add the vertex $N+1$, and join it with vertex $n$ in the appropriate manner. Now lets update vertex $N+1$ to employ its best response $r$. In other words, we shall let $N+1$ use the resource $r$ that maximizes its payoff, given the resource $X_n$ used by its new neighbor $n$.

There are two possibilities following this. The first possibility is that vertex $n$ remains satisfied (\ie $X_n$ is still $n$'s best response) after its new neighbor $N+1$ is added. In this case the state on $T$ is a Nash equilibrium of the new game $\psi$. The second possibility is that $X_n$ is no longer vertex $n$'s best response after the new vertex $N+1$ has been added, playing $r$. In this case we must have $X_n =r$. Now in this case, lets consider the modified game, restricted to the vertices $\{ 1 , 2 ,...,N \}$ of $T$, where vertex $n$'s payoff function $f^r _n$, for using resource $r$, is replaced with the payoff function $g^r _n$ such that $g^r _n (x) = f^r _n (x + S_{N+1,n})$, $\forall x$. Note that the payoff functions associated with all other players/resources within this modified game are the same as those within $\psi$. Now this modified game (which, we consider to be played upon $T'$, whilst temporally ignoring vertex $N+1$) is another GCGWE on an $N$ vertex quasi-tree. It follows (by assumption) that this modified game has a Nash equilibrium. Now suppose we allocate strategies to the vertices of $T'$ in accordance with such a Nash equilibrium ${ \boldsymbol Y}$. Let $Y_m$ denote the resource allocated to player $m \in \{ 1 , 2 ,..., N \}$ within this Nash equilibrium of the modified game upon $T'$. Now let us define the state ${ \boldsymbol Z}$ of game $\psi$ such that $Z_m = Y_m$, $\forall m \in \{ 1 , 2 ,..., N \}$ and $Z_{N+1} = r$. To see that ${ \boldsymbol Z}$ is a Nash equilibrium of $\psi$ note that there are two possibilities.
\begin{itemize}
\item The first possibility is that $Z_n = Z_{N+1} = r$. In this case each player in $\{ 1 , 2 ,..., N \}$, including $n$, is playing the best responses to their surroundings. Also, when $N+1$ was previously added to the game, we found that $N+1$'s best response was $r$, even though $N+1$'s only neighbor was playing $r$. In state ${ \boldsymbol Z}$ we also have that $N+1$ is using the same resource, $r$, as its neighbor $n$. It follows that $N+1$ is playing its best response in the configuration ${\boldsymbol Z}$. This implies that ${\boldsymbol Z}$ is  a Nash equilibrium.

\item The second possibility is that $Z_n \neq Z_{N+1} = r$.
Now in this case each player in $\{ 1 , 2 ,..., N \}$, including $n$, is playing the best responses to their surroundings.
Also, when $N+1$ was previously added to the game, we found that $N+1$'s best response was $r$, even though $N+1$'s only neighbor was playing $r$, and $N+1$ was suffering congestion from $n$. In state ${ \boldsymbol Z}$, player $N+1$ is using resource $r$, but now it does not suffer any congestion from its neighbor $n$ for doing so. It follows that $N+1$ is playing its best response in the configuration ${\boldsymbol Z}$. This implies that ${\boldsymbol Z}$ a Nash equilibrium.
\end{itemize}
So we have shown that the game $\psi$, on the $N+1$ vertex quasi-tree has a Nash equilibrium. Now our argument implies that if every GCGWE on an $N$ vertex quasi-tree has a Nash equilibrium then every game on an $N+1$ vertex quasi-tree has a Nash equilibrium.
This completes the induction proof.
\hfill $\Box$

 \subsection{Proof of Theorem \ref{dagcomplex}}\label{appendix:proofTheroem2}

 The key observation is that every directed acyclic graph can be given a \emph{topological sort} \cite{topology}. A topological sort is an ordering of the vertices, such that if $i < j$ then there is no directed edge from the $j$th vertex to the $i$th vertex in the ordering. Intuitively, a topological sort is a way to arrange the vertices of a directed acyclic graph in a line, so that every directed edge ``points towards the right''.
The best response of a player $n$, in state ${ \boldsymbol X}$, is the strategy $r \in \mathcal{R}_n$ which maximizes $n$'s payoff, given the strategies of the other players $m$ within ${ \boldsymbol X}$.
Consider a GCGWE on a directed acyclic graph $D(S)$ with adjacency matrix $S$. Suppose we select some topological sort of $D(S)$.  Let $k(u) \in \mathcal{N}$ denote the $u$th vertex/player which appears in the topological sort of $D(S)$. Now we can construct a Nash equilibrium of the GCGWE by having players sequentially update their strategies according to their best responses in the order of $k(1),k(2),k(3), \cdots$. In other words, player $k(1)$ updates to its best response, then  player $k(2)$ updates to its best response, so on and so forth,  until all players have been updated.
This will lead to a Nash equilibrium, because player $k(j)$'s best response update will not affect players $k(i)$ with $ i < j$, who have already updated. In other words, no player will ever regret their decision, because players updated subsequently will have no influence upon them. \hfill $\Box$

\subsection{Proof of Theorem \ref{twor}}\label{appendix:prooftheorem3}

Suppose we have a GCGWE with resource set $\mathcal{R} = \{ 1 , 2 \}$ and a symmetric spatial matrix $S$. For each player $n \in \mathcal{N}$, we can define a parameter $T_n$,  which corresponds to the maximal amount of congestion that player $n$ can tolerate from neighbors using resource $2$, before it prefers to change to resource $1$.
\begin{itemize}
\item If $f_n ^1 (x) <  f_n ^2 \left( \left( \sum_{m = 1} ^N S_{m,n} \right) - x \right) $ for all $ x \in \left[0, \sum_{m = 1} ^N S_{m,n} \right]$, then $T_n = 1 + \sum_{m = 1} ^N S_{m,n}$. In this case $n$ will always prefer resource $2$ to resource $1$, no matter how congested resource $2$ is.
\item  If $f_n ^1 (x) > f_n ^2 \left( \left( \sum_{m = 1} ^N S_{m,n} \right) - x \right)$ for all $x \in \left[0, \sum_{m = 1} ^N S_{m,n} \right]$, then $T_n = -1$. In this case $n$ will always prefer resource $1$ to resource $2$, even if resource $2$ is not congested.
\item Otherwise there must be a unique $x^*$ in $ \left[0, \sum_{m = 1} ^N S_{m,n} \right]$ such that $f_n ^1 (x^*) = f_n ^2 \left( \left( \sum_{m = 1} ^N S_{m,n} \right) - x^* \right)$ (since each $f_n^r$ is continuous and decreasing).  In this case $T_n = \left( \sum_{m = 1} ^N S_{m,n} \right) - x^*$.
\end{itemize}

\rev{The rest of the proof is based on constructing a potential function, which is a mapping from each state to a real number and decreases with each better response update.\footnote{The usual definition of potential function requires the function to increase with the best response update. Here we just inverse the sign and construct a function that monotonically decreases instead.} This implies that it is impossible for the system to visit the same state more than once. Since the total number of states is finite, this implies that every sufficiently long sequence of better response updates reaches a state from which no further better response updates can be performed. Such a state is a Nash equilibrium by definition.


Next we define the potential $V$ as a function for any state ${ \boldsymbol X}$,
\begin{equation}\label{startingeqn}
V({ \boldsymbol X}) = \frac{1}{2}\left( \sum_{m=1} ^N \sum_{m'=1} ^N S_{m',m} (X_m - 1 ) ( X_{m'} - 1) \right) - \sum_{m=1} ^N T_m (X_m - 1).
\end{equation}
We will show that $V$ decreases with each better response update.

We begin by defining
\begin{equation}\label{bbb}
A({ \boldsymbol X}) = \frac{1}{2}\left( \sum_{m=1} ^N \sum_{m'=1} ^N S_{m',m} (X_m - 1 ) ( X_{m'} - 1) \right)
\end{equation}
and
\begin{equation}\label{bbbb}
B({ \boldsymbol X}) = \sum_{m=1} ^N T_m (X_m - 1).
\end{equation}
Clearly  $V({ \boldsymbol X}) = A({ \boldsymbol X}) - B({ \boldsymbol X})$. Now we will show that
\begin{equation}\label{expasms}
A({\boldsymbol X}) = \frac{1}{2}\left( \sum_{m\neq n} \sum_{m' \neq n }  S_{m',m} (X_m - 1 ) ( X_{m'} - 1)   \right) + ( X_{n} - 1)\left( \sum_{m \neq n }  S_{n,m} (X_m - 1 )  \right).
\end{equation}
To see (\ref{expasms}), first note that we can expand out the right hand side of (\ref{bbb}) to obtain
\begin{equation}\label{expa}
\begin{split}
A({\boldsymbol X}) =& \frac{1}{2}\left( \sum_{m\neq n} \left( \sum_{m' \neq n }  S_{m',m} (X_m - 1 ) ( X_{m'} - 1) \right) + S_{n,m} (X_m - 1 ) ( X_{n} - 1)  \right) \\
  & + \frac{1}{2}\left( \sum_{m' \neq n }  S_{m',n} (X_n - 1 ) ( X_{m'} - 1) \right) + S_{n,n} (X_n - 1 ) ( X_{n} - 1).
\end{split}
\end{equation}
We can further use the fact $S_{n,n}=0$ to rewrite (\ref{expa}) as
\begin{equation}\label{expas}
\begin{split}
A({\boldsymbol X}) =& \frac{1}{2}\left( \sum_{m\neq n} \sum_{m' \neq n }  S_{m',m} (X_m - 1 ) ( X_{m'} - 1)   \right) + \frac{1}{2}\left( \sum_{m \neq n }  S_{n,m} (X_m - 1 ) ( X_{n} - 1) \right) \\
  & + \frac{1}{2}\left( \sum_{m' \neq n }  S_{m',n} (X_n - 1 ) ( X_{m'} - 1) \right).
\end{split}
\end{equation}
Now by changing the dummy variable in the third part of the right hand side of (\ref{expas}) from $m'$ to $m$, we obtain
\begin{equation}\label{expasm}
\begin{split}
A({\boldsymbol X}) =& \frac{1}{2}\left( \sum_{m\neq n} \sum_{m' \neq n }  S_{m',m} (X_m - 1 ) ( X_{m'} - 1)   \right) + \frac{1}{2}\left( \sum_{m \neq n }  S_{n,m} (X_m - 1 ) ( X_{n} - 1) \right) \\
  & + \frac{1}{2}\left( \sum_{m \neq n }  S_{m,n} (X_n - 1 ) ( X_{m} - 1) \right).
\end{split}
\end{equation}
Since $S_{m,n} = S_{n,m}$ holds on an undirected graph, we can simplify (\ref{expasm}) to obtain (\ref{expasms}).

Suppose that our game moves from state ${ \boldsymbol X}$ to state ${ \boldsymbol Y}$ because a player $n$ performs a better response update. We will now show that
\begin{equation}\label{maineqn}
V ( { \boldsymbol Y} ) = V ( { \boldsymbol X} ) + (Y_n - X_n) \left( \left( \sum_{m=1} ^N S_{m,n} ( X_m - 1 ) \right) - T_n \right).
\end{equation}
To see this, firstly note that in a similar way of deriving (\ref{expasms}) we can obtain
\begin{equation}\label{yexpasms}
A({\boldsymbol Y}) = \frac{1}{2}\left( \sum_{m\neq n} \sum_{m' \neq n }  S_{m',m} (Y_m - 1 ) ( Y_{m'} - 1)   \right) + ( Y_{n} - 1)\left( \sum_{m \neq n }  S_{n,m} (Y_m - 1 )  \right).
\end{equation}
Now, since $m \neq n$ implies $Y_m = X_m$, we can rewrite (\ref{yexpasms}) as
\begin{equation}\label{yexpasmss}
A({\boldsymbol Y}) = \frac{1}{2}\left( \sum_{m\neq n} \sum_{m' \neq n }  S_{m',m} (X_m - 1 ) ( X_{m'} - 1)   \right) + ( Y_{n} - 1)\left( \sum_{m \neq n }  S_{n,m} (X_m - 1 )  \right).
\end{equation}
Now subtracting (\ref{expasms}) from (\ref{yexpasmss}) yields
\begin{equation}\label{point}
A({\boldsymbol Y}) - A({\boldsymbol X}) =  ( Y_{n} - 1)\left( \sum_{m \neq n }  S_{n,m} (X_m - 1 )  \right) - ( X_{n} - 1)\left( \sum_{m \neq n }  S_{n,m} (X_m - 1 )  \right).
\end{equation}
We can rearrange (\ref{point}) to obtain
\begin{equation}\label{point2}
A({\boldsymbol Y}) = A({\boldsymbol X}) + ( Y_{n} - X_n)\left( \sum_{m \neq n }  S_{n,m} (X_m - 1 )  \right).
\end{equation}
Also note that
\begin{equation}\label{bbbbc}
B({ \boldsymbol Y}) = \sum_{m=1} ^N T_m (Y_m - 1) = \left(\sum_{m=1} ^N T_m (X_m - 1) \right) + T_n (Y_n - X_n)= B({ \boldsymbol X}) + T_n (Y_n - X_n),
\end{equation}
where the last equation  of (\ref{bbbbc}) is obtained by substituting (\ref{bbbb}). Now subtracting (\ref{bbbbc}) from (\ref{point2}) yields
\begin{equation}\label{sdone}
\begin{split}
V({ \boldsymbol Y}) & = A({ \boldsymbol Y}) - B({ \boldsymbol Y}) \\
& = A({\boldsymbol X}) + ( Y_{n} - X_n)\left( \sum_{m \neq n }  S_{n,m} (X_m - 1 ) \right) - B({ \boldsymbol X}) - T_n (Y_n - X_n) \\
& = V ( { \boldsymbol X} ) + (Y_n - X_n) \left( \left( \sum_{m \neq n} S_{n,m} ( X_m - 1 ) \right) - T_n \right).
\end{split}
\end{equation}
Since $S_{m,m} = 0$ and $S_{n,m} = S_{n,m}$, we have
\begin{equation}\label{pqp}
 \sum_{m \neq n} S_{n,m} ( X_m - 1 )  =  \sum_{m =1} ^N S_{m,n} ( X_m - 1 ).
\end{equation}
Finally we can substitute (\ref{pqp}) into (\ref{sdone}) to obtain (\ref{maineqn}), as required.

Next we will show that $V ( { \boldsymbol Y} ) <  V ( { \boldsymbol X} )$. Since state ${ \boldsymbol Y}$ comes from state ${ \boldsymbol X}$  after having player $n$ perform a better response update, there are two possibilities:
\begin{enumerate}
\item[(a)]  $X_n = 1$, $Y_n = 2$, and $f_n ^2 \left( \sum_{m:X_m = 2} S_{m,n} \right) > f_n ^1 \left( \sum_{m:X_m = 1} S_{m,n} \right)$; or
\item[(b)] $X_n = 2$, $Y_n = 1$, and $f_n ^2 \left( \sum_{m:X_m = 2} S_{m,n} \right) < f_n ^1 \left( \sum_{m:X_m = 1} S_{m,n} \right)$.
 \end{enumerate}
Under case (a), we  have $(Y_n - X_n) >0$ and $\sum_{m:X_m = 2} S_{m,n} = \sum_{m=1}^N S_{m,n} (X_m -1) < T_n$, thus based on (\ref{maineqn}) we have $V ( { \boldsymbol Y} ) < V ( { \boldsymbol X} )$. Under case (b),  we  have $(Y_n - X_n) <0$ and $\sum_{m:X_m = 2} S_{m,n} = \sum_{m=1}^N S_{m,n} (X_m -1) > T_n$, thus based on (\ref{maineqn}) we have  $V ( { \boldsymbol Y} ) < V ( { \boldsymbol X} )$. This completes the proof. \hfill $\Box$

}

\subsection{A Lemma for the Proof of Theorem \ref{sym}}\label{appendix:lemma1}


Let us define the total congestion level of a state ${ \boldsymbol X}$ to be the sum of the congestion levels of all players, \ie $C( { \boldsymbol X}) = \sum_{n = 1 } ^N \sum_{m : X_n = X_{m}} S_{m,n}$ . Lemma \ref{symcool} states that $C({ \boldsymbol X})$ decreases every time a player performs a better response update.
\begin{lemma}\label{symcool}
Suppose we have a GCGWE with homogenous resources and a symmetric spatial matrix $S$, that starts in state ${ \boldsymbol X}$. Now suppose that some player performs a better response update, and this converts the game state to $\boldsymbol{Y}$. We have $C({ \boldsymbol Y}) < C({ \boldsymbol X})$.
\end{lemma}

\emph{Proof:} \rev{Let's use $c _{n} ( { \boldsymbol X}) =  \sum_{m : X_{m} = X_n} S_{m,n}$ to denote the congestion level of a player $n$ when the system is in a state ${ \boldsymbol X}$. Suppose the system is in state ${\boldsymbol X}$, and then player $n$ performs a better response update, by changing his resource choice from $X_n$ to $r$. Suppose ${ \boldsymbol Y}$ is the state that results from this update. Since this update is a better response, Theorem \ref{simp} implies
\begin{equation}\label{ss0}
c _{n} ( { \boldsymbol Y}) = \sum_{m \in \mathcal{N}: X_{m} = r} S_{m,n}< \sum_{m \in \mathcal{N}: X_{m} = X_n} S_{m,n} = c _{n} ( { \boldsymbol X}).
\end{equation}
Clearly $n$'s congestion level decreases by $c _{n} ( { \boldsymbol X}) - c _{n} ( { \boldsymbol Y})$ as a result of his better response update.

Next we show that the sum of the congestion levels of $n$'s neighbors also decreases by $c _{n} ( { \boldsymbol X}) - c _{n} ( { \boldsymbol Y})$ as a result of $n$'s better response update.
Since  $C ( { \boldsymbol Z} ) = \sum_{n=1} ^N \sum_{m: Z_{m} = Z_n} S_{m,n} = \sum_{n=1} ^N c _{n} ( { \boldsymbol Z})$, then we must have $C( { \boldsymbol X} ) - C( { \boldsymbol Y} ) = 2(c _{n} ( { \boldsymbol X}) - c _{n} ( { \boldsymbol Y}))>0$ (including the congestion level decrease of player $n$ and his neighbors).

The proof relies on the following three statements for a generic player $m\neq n$.
\begin{enumerate}
\item If a player $m$ does not use either resource $X_{n}$ or resource $r$ in state $\boldsymbol{X}$, then his congestion level will not change in state state $\boldsymbol{Y}$.
\item If player $m$ uses resource $X_{m} = X_n$ in state ${ \boldsymbol X}$, then $c _{m} ( { \boldsymbol Y}) = c _{m} ( { \boldsymbol X}) - S_{n,m} = c _{m} ( { \boldsymbol X}) - S_{m,n}$. This is due to the fact that $n$ chooses a different resource $r$ in state $\boldsymbol{Y}$ and no longer congests with $m$. Thus the congestion level of $m$ is reduced by $S_{n,m}$ (which is equal to $S_{m,n}$ because $S$ is symmetric).
\item If player $m $ uses resource  $X_{m} = r$ in state ${ \boldsymbol X}$,  then $c _{m} ( { \boldsymbol Y}) = c _{m} ( { \boldsymbol X}) + S_{n,m} = c _{m} ( { \boldsymbol X}) + S_{m,n}$. The is due to the fact that $n$ starts to congest with $m$ in state $\boldsymbol{Y}$.
\end{enumerate}
Statement 1 implies
\begin{equation}\label{ss1}
\sum_{m:X_m \notin \{X_n,r \}} c_m ({ \boldsymbol Y}) = \sum_{m:X_m \notin \{X_n,r \}} c_m ({ \boldsymbol X}).
\end{equation}
Statement 2 implies
\begin{equation}\label{ss2}
\sum_{m \neq n :X_m = X_n} c_m ({ \boldsymbol Y}) = \left(\sum_{m \neq n :X_m =X_n} c_m ({ \boldsymbol X})\right) - \sum_{m :X_m = X_n} S_{m,n}.
\end{equation}
Statement 3 implies
\begin{equation}\label{ss3}
\sum_{m : X_m = r} c_m ({ \boldsymbol Y}) = \left(\sum_{m :X_m =r } c_m ({ \boldsymbol X}) \right) + \sum_{m :X_m = r} S_{m,n}.
\end{equation}
Substituting the expression $c_n({ \boldsymbol X}) = \sum_{m :X_m = X_n} S_{m,n}$ into (\ref{ss2}) yields
\begin{equation}\label{sss2}
\sum_{m \neq n :X_m = X_n} c_m ({ \boldsymbol Y}) = \left(\sum_{m \neq n :X_m =X_n} c_m ({ \boldsymbol X})\right) - c_n({ \boldsymbol X})
\end{equation}
Substituting the expression $c_n({ \boldsymbol Y}) = \sum_{m :X_m = r} S_{m,n}$ into (\ref{ss3}) yields
\begin{equation}\label{sss3}
\sum_{m : X_m = r} c_m ({ \boldsymbol Y}) = \left(\sum_{m :X_m =r } c_m ({ \boldsymbol X})\right) + c_n({ \boldsymbol Y}).
\end{equation}
Also, obviously
\begin{equation}\label{obv}
c_n({ \boldsymbol Y}) = c_n({ \boldsymbol X}) - c_n({ \boldsymbol X}) + c_n({ \boldsymbol Y})
\end{equation}
Now we can write
\begin{equation}\label{allss}
C({\boldsymbol Y}) = \sum_{m \in \mathcal{N}} c_m ({ \boldsymbol Y}) = c_n({ \boldsymbol Y}) + \left(\sum_{m:X_m \notin \{X_n,r \}} c_m ({ \boldsymbol Y})\right) + \left( \sum_{m \neq n :X_m = X_n} c_m ({ \boldsymbol Y}) \right) + \left( \sum_{m : X_m = r} c_m ({ \boldsymbol Y}) \right)
\end{equation}
Now by using (\ref{obv}), (\ref{ss1}), (\ref{sss2}) and (\ref{sss3}) to substitute the respective terms on the right hand side of (\ref{allss}), we obtain
\begin{equation} \label{sresult}
\begin{split}
C({\boldsymbol Y}) =& c_n({ \boldsymbol X}) - c_n({ \boldsymbol X}) + c_n({ \boldsymbol Y}) + \left( \sum_{m \neq n :X_m = X_n} c_m ({ \boldsymbol X}) \right) + \left(\sum_{m \neq n :X_m =X_n} c_m ({ \boldsymbol X})\right) - c_n({ \boldsymbol X}) \\
  & + \left( \sum_{m :X_m =r } c_m ({ \boldsymbol X}) \right) + c_n({ \boldsymbol Y})
  \end{split}
\end{equation}
Next note that
\begin{equation}\label{ssimp}
C({\boldsymbol X}) = \sum_{m \in \mathcal{N}} c_m ({ \boldsymbol X}) = c_n({ \boldsymbol X}) + \left(\sum_{m:X_m \notin \{X_n,r \}} c_m ({ \boldsymbol X})\right) + \left( \sum_{m \neq n :X_m = X_n} c_m ({ \boldsymbol X}) \right) + \left( \sum_{m : X_m = r} c_m ({ \boldsymbol X}) \right)
\end{equation}
Now we can use (\ref{ssimp}) to simplify (\ref{sresult}) and obtain
\begin{equation} \label{ssresult}
 C({\boldsymbol Y}) =  C({\boldsymbol X}) + 2(c_n({ \boldsymbol Y}) - c_n({ \boldsymbol X}))
\end{equation}
Using (\ref{ssresult}) together with the fact that $c_n({ \boldsymbol X})>c_n({ \boldsymbol Y})$ gives us that $C({\boldsymbol X}) > C({\boldsymbol Y})$ as required. $\Box$}


This lemma also has an important corollary.

\begin{corollary}\label{bestconfig}
If ${ \boldsymbol X}$ is the state of a GCGWE with homogenous resources (played upon an undirected graph) that minimizes the total amount of congestion $C( { \boldsymbol X} )$, then ${ \boldsymbol X}$ will be a Nash equilibrium.
\end{corollary}

\emph{Proof:} We shall prove this result by contradiction. Suppose that ${ \boldsymbol X}$ is the state of our game which minimizes the total amount of congestion, and that ${ \boldsymbol X}$ is \emph{not} a Nash equilibrium. In this case, some better response can be performed from ${ \boldsymbol X}$, which moves the system into a different state, ${ \boldsymbol Y}$. However, Lemma \ref{symcool} implies that $C( { \boldsymbol Y} ) < C( { \boldsymbol X} ) $, which contradicts our assumption that state ${ \boldsymbol X}$ minimizes $C$. $\Box$

%

\subsection{Proof of Theorem \ref{nphard}}\label{appendix:proofTheroem7}

\rev{A problem $H$ is said to be NP hard, if there is an NP complete problem $L$ which can be solved in polynomial time given an oracle machine $O$ for $H$. Such an oracle machine $O$ is capable of solving any instance of $H$ in polynomial time.

Let $L$ denote the problem of determining if some graph $G$ can be given a proper coloring using three colors. A {proper coloring} of a undirected graph $G$ is an assignment of colors to the vertices of $G$ such that no two vertices with the same color are linked. $L$ is known to be NP complete (\cite{colorhard}). Let $H$ denote the problem of finding the Nash equilibrium of a GCGWE, on an undirected graph with homogenous resources, which maximizes the total payoff of the players. Now we will show that, given an oracle machine $O$ for $H$, we can solve $L$ in polynomial time. We will demonstrate this by showing that, for any undirected simple graph $G$, we can construct a GCGWE $g$ on $G$ (with homogenous resources) which has the following two properties:
\begin{itemize}
\item If $G$ can be given a proper coloring with three colors, then every pure Nash equilibrium ${\boldsymbol X}$ of $g$ that maximizes the total payoff of the players corresponds to a proper coloring of $G$ (in the sense that no two linked players are given the same resource under ${\boldsymbol X}$).
\item If $G$ can not be given a proper coloring with three colors, then every pure Nash equilibrium ${\boldsymbol X}$ of $g$ which maximizes the total payoff of the players does not correspond to a proper coloring of $G$ (in the sense that there exists some pair of linked players which are given the same resource under ${\boldsymbol X}$).
\end{itemize}

Suppose we have an instance of $L$. In other words, suppose that we have a simple graph $G = (V,E)$, and we wish to determine whether it can be given a proper coloring using three colors. Let $A$ be the adjacency matrix for the graph $G$. Now consider the GCGWE $g = ( \mathcal{N} ,\mathcal{R},(\mathcal{R}_n)_{n \in \mathcal{N}},(f_n ^r)_{n \in \mathcal{N},r \in \mathcal{R}_n},S),$ with a set of players $\mathcal{N} = V$, a set of three resources $\mathcal{R} = \{ 1 , 2 , 3 \} = \mathcal{R}_1 = \cdots =\mathcal{R}_N$ available to each player, a spatial matrix $S = A$ equal to the adjacency matrix of our graph, and payoff functions such that $f_n ^r (x) = -x$, $\forall n \in \mathcal{N}$, $\forall r \in \mathcal{R}, \forall x$. The GCGWE $g$ has three homogenous resources, and is played on the undirected graph $G$.

Now (by assumption) our oracle machine $O$ can find a Nash equilibrium ${ \boldsymbol X}$ of the GCGWE $g$ which maximizes the total payoff to the players, in polynomial time. Now Corollary \ref{bestconfig} in Appendix \ref{appendix:lemma1} implies that ${ \boldsymbol X}$ is one of the states of $g$ which minimizes the total congestion levels of the players in game $g$.

Now if $G$ can be given a proper coloring using only three colors, then the Nash equilibrium ${ \boldsymbol X}$ will clearly involve no congestion (because, it will be possible to allocate resources to the players in $g$ in such  that two  players $n$ and $m$ are given the same resource only if $S_{n,m} = A_{n,m} = 0$). Conversely, if $G$ cannot be given a proper coloring using only three colors, then it is inevitable that some pair of players will cause each other some congestion in any state ${ \boldsymbol X}$ of $g$.
%
Hence, given ${ \boldsymbol X}$ (which we can determine in polynomial time, using our oracle machine $O$) we can solve our instance of problem $L$ in order $n^2$ time by checking whether any pair of linked players $n,m \in \mathcal{N}$ are using the same resource. If we find a pair of linked players using the same resource, then the answer to our instance of $L$ is ``no'' (i.e., $G$ cannot be given a proper coloring using only three colors), otherwise the answer is ``yes'' (i.e., $G$ can be given a proper coloring using only three colors).

It follows that, if we have an oracle $O$, for $H$, then we have a polynomial time algorithm for solving $L$ (\ie $L$ is polynomial time reducible to $H$). Now since $L$ is NP complete, we have that $H$ is NP hard.} \hfill $\Box$


%





\bibliographystyle{nonumber}

\end{document}